\documentclass[12pt,aps,preprint,nofootinbib]{revtex4}

\usepackage{graphicx}
\usepackage{cancel}
\usepackage{amssymb}
\usepackage{textcomp}
\usepackage{amsmath}
\usepackage{bm}
\usepackage{times}
\usepackage{epsfig}
\usepackage{epstopdf}
\usepackage{color}
\usepackage{multirow}
\usepackage{subfig}
\usepackage[colorlinks=true, pdfstartview=FitV, linkcolor=blue, citecolor=blue, urlcolor=blue]{hyperref}

\def\lsim{\mathrel{\raise.3ex\hbox{$<$\kern-.75em\lower1ex\hbox{$\sim$}}}}
\def\gsim{\mathrel{\raise.3ex\hbox{$>$\kern-.75em\lower1ex\hbox{$\sim$}}}}

\def\beq{\begin{equation}}
\def\eeq{\end{equation}}
\def\be{\begin{equation}}
\def\ee{\end{equation}}
\def\bea{\begin{eqnarray}}
\def\eea{\end{eqnarray}}

\def\bsg{b\to s\gamma}
\def\Br{\rm BR}
\def\BR{\rm BR}

\def\ma{m_{A}}

\def\gev{\,{\rm GeV}}

\def\to{\rightarrow}

\begin{document}
%\title{Implication of LHC Searches on Co-annihilated Stop and Neutralino Dark Matter in Yukawa
%Unified CMSSM
\title{Decoupling MSSM Higgs Sector and Heavy Higgs Decay}
%\vspace{4mm}

\author{Tong Li}
%\email{than@pitt.edu}
%\author{M. Adeel Ajaib\footnote{email: adeel@udel.edu}}
%\author{Tong Li\footnote{email: tli@udel.edu, corresponding author}}
%\author{Qaisar Shafi\footnote{email: shafi@bartol.udel.edu}}
%\address{
%Bartol Research Institute, Department of Physics and Astronomy,
%University of Delaware, Newark, Delaware 19716, USA
%\vspace{3mm}
%}
%\date{\today}

\affiliation{ARC Centre of Excellence for Particle Physics at the
Terascale,
 School of Physics, Monash University,
 Melbourne, Victoria 3800, Australia
}

\begin{abstract}
The decoupling limit in the MSSM Higgs sector is the most likely scenario in light of the Higgs discovery.
This scenario is further constrained by MSSM Higgs search bounds and flavor observables. We perform a comprehensive scan of MSSM parameters and update the constraints on the decoupling MSSM Higgs sector in terms of 8 TeV LHC data. We highlight the effect of light SUSY spectrum in the heavy neutral Higgs decay in the decoupling limit. We find that the chargino and neutralino decay mode can reach at most 40\% and 20\% branching ratio, respectively. In particular, the invisible decay mode $BR(H^0(A^0)\to \tilde{\chi}^0_1\tilde{\chi}^0_1)$ increases with increasing Bino LSP mass and is between 10\%-15\% (20\%) for $30<m_{\tilde{\chi}^0_1}<100$ GeV. The leading branching fraction of heavy Higgses decay into sfermions can be as large as 80\% for $H^0\to \tilde{t}_1\tilde{t}_1^\ast$ and 60\% for $H^0/A^0\to \tilde{\tau}_1\tilde{\tau}_2^\ast+\tilde{\tau}_1^\ast\tilde{\tau}_2$. The branching fractions are less than 10\% for $H^0\to h^0h^0$ and 1\% for $A^0\to h^0Z$ for $m_A>400$ GeV. The charged Higgs decays to neutralino plus chargino and sfermions with branching ratio as large as 40\% and 60\%, respectively. Moreover, the exclusion limit of leading MSSM Higgs search channel, namely $gg,b\bar{b}\to H^0, A^0\to \tau^+\tau^-$, is extrapolated to 14 TeV LHC with high luminosities. It turns out that the $\tau\tau$ mode can essentially exclude regime with $\tan\beta>20$ for $L=300$ fb$^{-1}$ and $\tan\beta>15$ for $L=3000$ fb$^{-1}$.
\end{abstract}

\maketitle
%%%%%%%%%%%%%%%%%%%%%%%%%%%%%%%%%%%%%%%%%%%%%%%%%%%%%%%%
\section{Introduction}
%%%%%%%%%%%%%%%%%%%%%%%%%%%%%%%%%%%%%%%%%%%%%%%%%%%%%%%%%%%%

The discovery of the Higgs boson at the LHC~\cite{:2012gk} raises two questions to theoretical particle physicists about the Higgs mechanism: is the discovered Higgs boson a pure Standard Model (SM) Higgs or SM-like Higgs from new physics theory? can the LHC prove or disprove new physics associated with Higgs sector?
To answer these questions, it is important to investigate the implication of existing Higgs search data for extended Higgs sector in new physics framework and propose dedicated Higgs search signatures for experimentalists to test.

One of the best motivated theories beyond the SM is the weak scale supersymmetry (SUSY). In the framework of the Minimal Supersymmetric Standard Model (MSSM), unlike SM, the Higgs sector is composed of two Higgs doublets~\cite{Gunion:1989we,Djouadi:2005gj}. After electroweak symmetry breaking, one has five physical Higgses, namely two CP-even Higges $h^0, H^0$, one CP-odd one $A^0$ and charged Higgses $H^\pm$. Between the two CP-even Higgs bosons, the one which couples to gauge bosons more strongly is SM-like. Moreover, the tree level Higgs masses are only determined by CP-odd Higgs mass parameter $m_A$ and the ratio of two doublets' vacuum expectation values $\tan\beta$. Requiring the SM-like production cross sections of a Higgs boson of a 126 GeV mass with decay to diphoton and gauge bosons splits the MSSM Higgs parameters into two distinct regions~\cite{Christensen:2012ei}:
%It was elaborated \cite{Christensen:2012ei} that, for a Higgs boson of a 126 GeV mass, requiring the SM-like cross sections for diphoton and gauge bosons, the %MSSM Higgs parameters split into two distinct regions:
\begin{itemize}
\item[(a)] the ``non-decoupling" region with $m_A\lesssim130$~GeV and $\tan\beta<10$~\cite{Haber:1994mt}. In this region, the heavy CP-even state $H^0$ is SM-like, while the light CP-even Higgs $h^0$ and the CP-odd one $A^0$ are nearly degenerate in mass and close to $m_{Z}$, and the charged state $H^\pm$ is slightly heavier.
 \item[(b)] the ``decoupling'' region with $\ma \gsim 300$ GeV~\cite{Haber:1994mt}. In this region, the light CP-even Higgs $h^0$ is SM-like, while all the other Higgs bosons are nearly degenerate with $\ma$ \cite{Haber:1995be}.
\end{itemize}
The non-decoupling scenario leads to light non SM-like Higgs states which could be searched immediately without SUSY parameter dependence~\cite{Christensen:2012si}.
%could be of immediate relevance for the LHC phenomenology,
However, this scenario is highly constrained by both MSSM Higgs search bounds and $b$-quark rare decays~\cite{nondecoupling}. The decoupling limit could thus be the most likely MSSM Higgs scenario in light of MSSM Higgs search results and the measurements of low-energy observables.

The leading channels probing decoupling scenario are the production of heavy neutral Higgses $H^0,A^0$ from gluon fusion, $b\bar{b}$ annihilation and associated process with $b$ quarks in final state, followed by decay into $b\bar{b}$ or $\tau^+\tau^-$~\cite{Arbey}. In particular, with tau Yukawa coupling enhanced in large $\tan\beta$ regime, the $\tau\tau$ decay mode puts the most stringent constraints on the heavy Higgs states as the $b\bar{b}$ production would be overwhelmed by a huge QCD background. However, the current bound and exclusion limit of $\tau\tau$ channel are generally based on predictions from generic two Higgs doublet model or some particular SUSY benchmarks~\cite{CMStau}. As well known, the fit to 126 GeV Higgs mass and signal excesses leads to light SUSY sparticles, for instance superpartners of top quark and tau lepton. Given light SUSY spectrum, the heavy neutral Higgses decay would change dramatically and result into altered exclusion limit of $\tau\tau$ channel~\cite{Ian}. The SUSY products effect in the heavy Higgs decay would also open rich LHC phenomenology. This paper aims to examine the current status of decoupling scenario and future perspectives for heavy Higgses decay and production. We highlight the complex pattern of heavy Higgses decay, in particular for small $\tan\beta$ region, taking into account the updated Higgs search bounds and latest flavor measurements. We perform the
extrapolation of $\tau\tau$ mode to the center-of-mass energy of 14 TeV with high luminosities at the LHC.

The rest of the paper is organized as follows. In Sec. II, we present the parameter choices relevant for Higgs observation in our scan. We also present the scanning results with subject to the constraints from the searches of Higgs and sparticles and flavor measurements. We also highlight the exotic patterns of heavy Higgs decay and extrapolate the $\tau\tau$ decay mode in Sec. III. We summarize our results in Sec. IV.

%%%%%%%%%%%%%%%%%%%%%%%%%%%%%%%%%%%%%%%%%%%%%%%%%%%%%%%
\section{SUSY Parameter region and experimental bounds}
%%%%%%%%%%%%%%%%%%%%%%%%%%%%%%%%%%%%%%%%%%%%%%%%%%%%%%%%%%%

To figure out the impact of experimental data on SUSY, it is crucial to scan the parameters relevant for the current Higgs observation and flavor measurements and extract the surviving space. We follow the procedure in Ref.~\cite{Christensen:2012ei} to explore the consistent parameter space.
To perform a comprehensive scan over the MSSM parameter space, besides the parameters adopted in Ref.~\cite{Christensen:2012ei}, we take into account the stau sector in the scan
%To explore the parameter space which is consistent with the current Higgs
%observation including the flavor constraints, we follow the
%procedure as in Ref.~\cite{Christensen:2012ei}, and perform a
%comprehensive scan over the MSSM parameter space
\begin{eqnarray}
 & 1 < \tan\beta < 55, \quad
 50\ {\rm GeV}< M_A < 1000\ {\rm GeV}, \quad 100\gev <  \mu < 2000\ {\rm GeV}, &\\
&   100\gev < M_{\tilde{t}_R}, M_{\tilde{Q}_3} < 2000\ {\rm GeV},
\quad
    {-4000\ \gev}< A_t < 4000\  {\rm GeV}, &\\
&    100\gev < M_{\tilde{\tau}_R}, M_{\tilde{L}_3} < 2000\ {\rm
GeV}, \quad
    {-4000\ \gev}< A_\tau < 4000\  {\rm GeV}, &\\
& 100 \ {\rm GeV}< M_2 < 2000\ {\rm GeV}. &
 \label{eq:para}
\end{eqnarray}
%with $X_t=A_t-\mu/\tan\beta$ and $X_\tau=A_\tau-\mu\tan\beta$.
In addition, we focus on the reduced high $M_A$ range in order to study the decoupling region:
\begin{equation}
 300\ {\rm GeV}< M_A < 1000\ {\rm GeV}.%  , \quad 0<A_t<4000\ {\rm GeV}.
 \label{eq:ma}
\end{equation}
%The $SU(2)_L$ gaugino mass parameter has significant impact on the
%flavor sector in the non-decoupling region and we scan it over a
%low-value range
%\begin{eqnarray}
%100 \ {\rm GeV}< M_2 < 300\ {\rm GeV}.
%\end{eqnarray}
The $U(1)$ gaugino mass $M_1$, however, is unconstrained in the MSSM
since Bino does not contribute much to either the Higgs sector, or the flavor observables.
Moreover, as indicated by the measurement of dark matter relic density, the dark matter candidate in the MSSM is more likely to
be a Bino-like neutralino with a mass heavier than 30 GeV~\cite{carlos,taodm}.
We thus prefer the Bino neutralino as the lightest supersymmetric particle (LSP) and take $m_{\tilde{\chi}_1^0}\approx M_1=90$ GeV for illustration, unless stated otherwise.
%without loss of generality
%\begin{eqnarray}
%M_1=90 \ {\rm GeV}.
%\end{eqnarray}
Other SUSY soft masses, which are less relevant to our
consideration, are all fixed to be 3 TeV.

%%%%%%%%%%%%%%%%%%%%%%%%%%

\subsection{Constraints from the Higgs Searches and $b$ Rare Decays}

We perform our scan by using the FeynHiggs 2.9.5
package~\cite{Degrassi:2002fi,Heinemeyer:1998np,Frank:2006yh,Heinemeyer:1998yj}
to calculate the Higgs masses, SUSY spectrum, couplings and Higgs decay/production rates.
HiggsBound 4.0.0 \cite{Bechtle:2008jh} is used to impose the exclusion
constraints from LEP2~\cite{LEP2H}, the Tevatron~\cite{CDFD0} and
the LHC.
%~\cite{ATLASrr,ATLASww,ATLASzz,CMSrr,CMSww,CMSzz,ATLASvhbb,ATLAStthbb,CMSvhbb,CMStthbb,ATLAStata,ATLAStataMSSM,CMSmumuMSSM,CMSbbMSSM,ATLASHpmta,CMSHpmta,ATLASHpmcs}.
%We generate a large random data sample that passes these
%constraints.
We further require that the light CP-even Higgs boson is
SM-like and satisfies the following properties
\begin{eqnarray}
&&h^0 \ {\rm in \ the \ mass \ range \ of} \ 124 \ {\rm GeV} - 128 \ {\rm GeV},\\
&&  \sigma \times {\BR} (gg\to h^0   \to\gamma\gamma)_{\rm
MSSM}\geq 80\% (\sigma\times {\BR})_{\rm SM},\\
&&   \sigma \times {\BR} (gg\to h^0   \to WW/ZZ)_{\rm MSSM}\geq 40\%
(\sigma\times {\BR})_{\rm SM}.
\end{eqnarray}

%\subsection{Constraints from $b$ Rare Decays}

The experimental flavor measurements considered here include
$\bsg$ \cite{Amhis:2012bh} and the LHCb report on $B_s\to \mu^+
\mu^-$ \cite{Aaij:2012nna}. In our study, we use the following experimental limits
\begin{eqnarray}
{\Br}(B_s \rightarrow X_s \gamma)_{\rm exp} = (3.43\pm 0.21)\times
10^{-4}, \ \
{\Br}(B_s\to \mu^+ \mu^-)_{\rm exp} = (2.9^{+1.1}_{-1.0})\times
10^{-9},
\end{eqnarray}
which are consistent with SM predictions~\cite{Misiak:2006zs,Misiak:2006ab,bsmumuSM}
\begin{eqnarray}
{\Br}(B_s \rightarrow X_s \gamma)_{\rm SM} = (3.15\pm 0.23)\times 10^{-4}, \ \ {\Br}(B_s\to \mu^+ \mu^-)_{\rm SM} = (3.23\pm 0.27)\times
10^{-9}.
\end{eqnarray}
BABAR also reported improved measurements of $B\to D \tau \nu_\tau$
which indicates a deviation from the SM expectation.
% and is sensitive
%to new physics contributions in the form of a light charged Higgs
%boson at tree level.
We take the observed excess as an upper
limit~\cite{Lees:2012xj}
\begin{eqnarray}
&& {{ {\Br}(B\to D\tau\nu_\tau) } \over { {\Br}(B\to
D\ell\nu_\ell)}} <0.44 , \quad
{{ {\Br}(B\to D\tau\nu_\tau)_{\rm SM}} \over { {\Br}(B\to D\ell\nu_\ell)_{\rm SM} }} =0.297\pm 0.017.% , \\
%&& {{ {\Br}(B\to D^{\ast}\tau\nu_\tau) } \over { {\Br}(B\to
%D^{\ast}\ell\nu_\ell)}} < 0.332, \ \ {{ {\Br}(B\to
%D^{\ast}\tau\nu_\tau)_{\rm SM}} \over { {\Br}(B\to
%D^{\ast}\ell\nu_\ell)_{\rm SM} }} = 0.252\pm 0.003.
\end{eqnarray}
In our numerical study, we use SuperIso 3.3~\cite{superiso} to
evaluate the above flavor observables.
%We also check that the values
%of ${\Br}(B\to \tau\nu)$, $\Delta m_{B_d}$ and $\Delta m_{B_s}$ in
%the non-decoupling region are consistent with the recent
%Belle~\cite{btaunu} and LHCb~\cite{deltam} measurements,
%respectively. As discussed in the previous section, the $\bsg$
%process gives the most stringent constraint on non-decoupling region
%with light charged Higgs. To achieve the SM-like measurement, in
%MSSM, the dominant charged Higgs-top loop contribution should be
%largely cancelled by other SUSY loops, in particular, the
%chargino-stop loop, which requires small values of $M_2$ and
%$M_{\tilde{Q}_3}$.

%%%%%%%%%%%%%%%%%%%%%%%%%

\subsection{Results for Allowed Region}

\begin{figure}[tb]
\begin{center}
\includegraphics[scale=1,width=8cm]{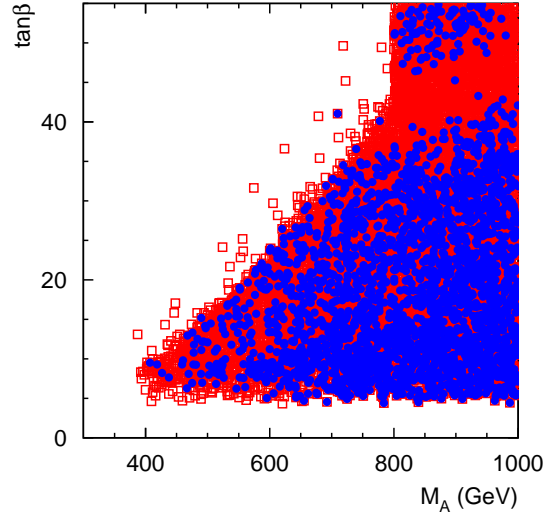}
\end{center}
\caption{$\tan\beta$ vs. $m_A$ for surviving points satisfying bounds from LEP2, Tevatron, LHC and $m_{h^0}=126\pm 2$ GeV (red open square), and further including $b$ rare decay constraints (blue filled circle). The following figures are all for points passing all constraints considered here.} \label{tbma}
\end{figure}

We generate sufficient random data samples and pass them through the above constraints.
Taking into account both the Higgs search results and the flavor constraints, we first show the
surviving points in Fig.~\ref{tbma} in the $\tan\beta-m_A$ plane. One can see that the measured Higgs mass window and current Higgs search data push the lower limit of $m_A$ to 400 GeV. Further $b$ rare decay constraints allow the whole region of $m_A>400$ GeV and $5<\tan\beta<40$. However, due to the enhancement of MSSM contributions to $B_s\to \mu^+\mu^-$ by $\tan^6\beta$ and reduction by $1/m_A^4$, the large $\tan\beta$ and small $m_A$ regime is highly constrained by $b$ rare decays. Note that although some points have $\tan\beta\gtrsim 45$, more data probing for heavy Higgs regime in near future would immediately restrict $m_A>800$ GeV with large $\tan\beta$. In the following we examine the surviving region favored by Higgs observation and flavor constraints.

\begin{figure}[tb]
\begin{center}
\subfloat[\label{dd:a}]{
\includegraphics[scale=1,width=7.5cm]{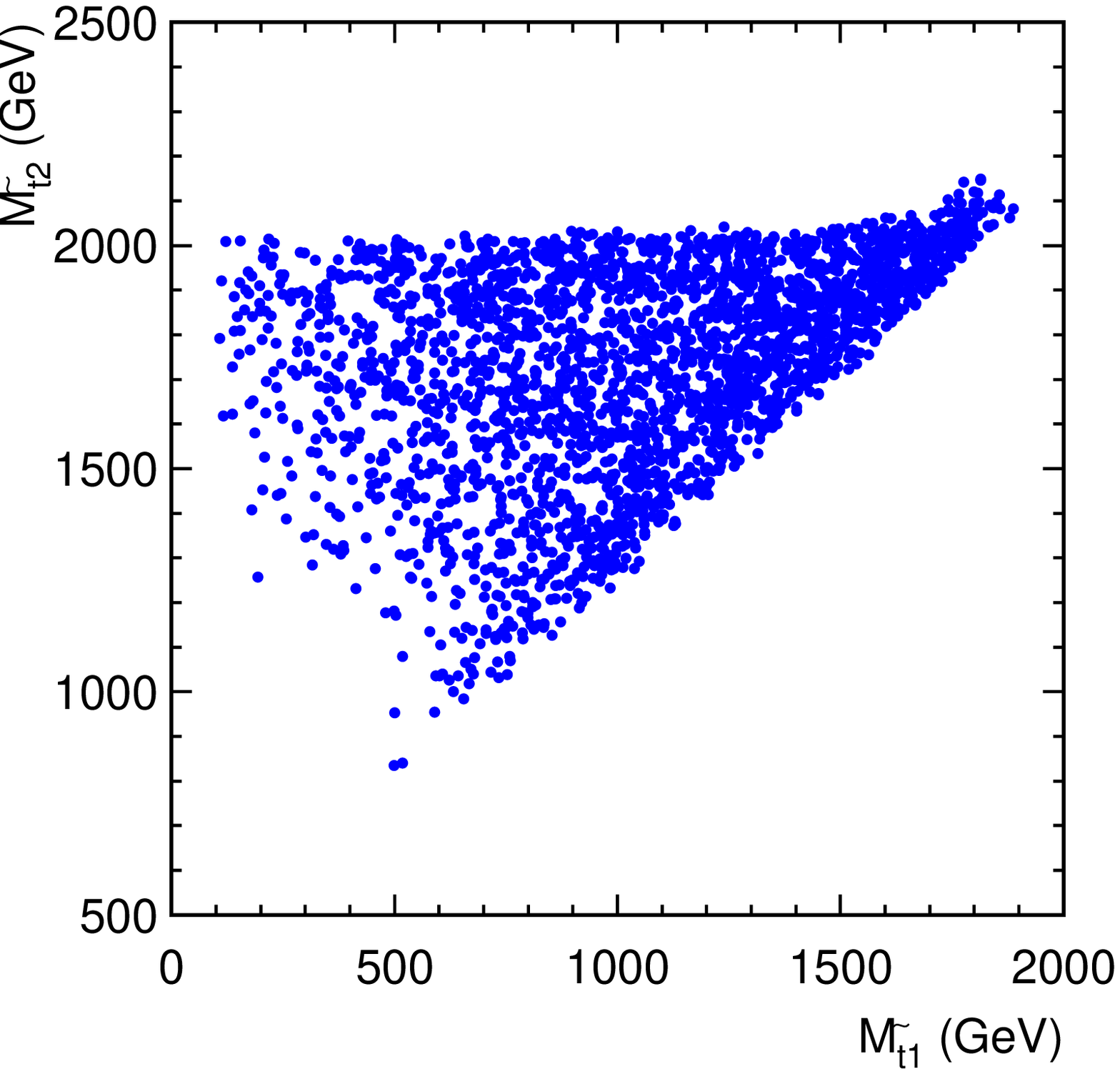}}
\subfloat[\label{dd:b}]{
\includegraphics[scale=1,width=7.5cm]{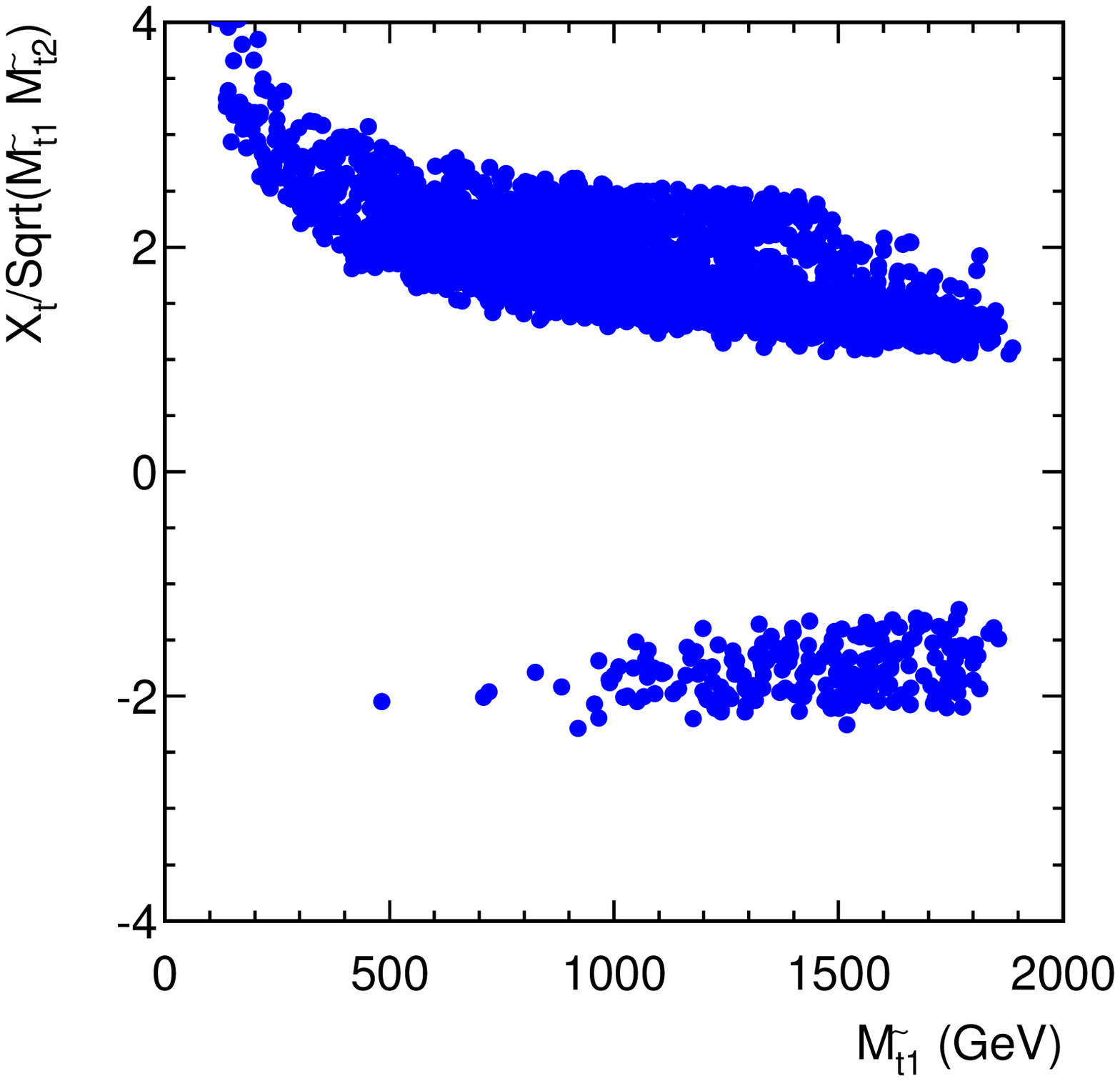}}
\end{center}
\caption{(a) $m_{\tilde{t}_2}$ vs. $m_{\tilde{t}_1}$ and (b) $X_t/\sqrt{m_{\tilde{t}_1}m_{\tilde{t}_2}}$ vs. $m_{\tilde{t}_1}$.} \label{stop}
\end{figure}

In the MSSM, as is well-known, the loop correction of the lightest MSSM Higgs mass is dominated by the stop sector and can raise $m_{h^0}$ to the observed value of Higgs boson mass. The leading stop loop correction is given by~\cite{Carena:1995wu}
\begin{eqnarray}
\epsilon = {3m_t^4\over 2\pi^2v^2 \sin^2\beta} \left[{\rm ln}\left({M_S^2\over m_t^2}\right)+{X_t^2\over M_S^2}\left(1-{X_t^2\over 12M_S^2}\right)\right],
\end{eqnarray}
where $X_t=A_t-\mu \cot\beta$ and $M_S=\sqrt{m_{\tilde{t}_1}m_{\tilde{t}_2}}$. Thus, as the measured Higgs mass is relatively heavier than tree level MSSM Higgs, the stop masses and stop mixing parameter, $X_t$, are strongly related to the Higgs mass in the MSSM. Satisfying the Higgs mass constraint, the stop masses are approximately given by~\cite{carlosstopstau}
\begin{eqnarray}
m_{\tilde{t}_1}^2\simeq m_{\tilde{Q}_3}^2+m_t^2\left(1-{X_t^2\over m_{\tilde{t}_R}^2}\right), \ m_{\tilde{t}_2}^2\simeq m_{\tilde{t}_R}^2+m_t^2\left(1+{X_t^2\over m_{\tilde{t}_R}^2}\right), \ {\rm for} \ |X_t|\simeq m_{\tilde{t}_R}\gg m_{\tilde{Q}_3},
\label{stopmass}
\end{eqnarray}
with the switch of $m_{\tilde{Q}_3}\leftrightarrow m_{\tilde{t}_R}$ for $|X_t|\simeq m_{\tilde{Q}_3}\gg m_{\tilde{t}_R}$, unless both stops are very heavy. The light stop is thus mostly left-handed (right-handed) and its mass is governed by $m_{\tilde{Q}_3}$ ($m_{\tilde{t}_R}$) for $m_{\tilde{t}_R}\gg m_{\tilde{Q}_3}$ ($m_{\tilde{Q}_3}\gg m_{\tilde{t}_R}$). The physical stop masses are
shown in Fig.~\ref{stop} (a).
%As can be seen, the light stop mass can be lighter than top quark.
As seen from the stop mixing effect in Fig.~\ref{stop} (b) in the plane of $X_t/\sqrt{m_{\tilde{Q}_3}m_{\tilde{t}_R}}$ vs. $m_{\tilde{t}_1}$, the ranges of
$X_t, m_{\tilde{Q}_3}, m_{\tilde{t}_R}$ sit nearly maximal stop mixing for light stops. Note that the values of light sbottom and sneutrino mass are determined by $m_{\tilde{Q}_3}$ and $m_{\tilde{L}_3}$, respectively, and thus mostly $\tilde{b}_L$ and $\tilde{\nu}_{\tau L}$.

\begin{figure}[tb]
\begin{center}
\subfloat[\label{dd:a}]{
\includegraphics[scale=1,width=7.5cm]{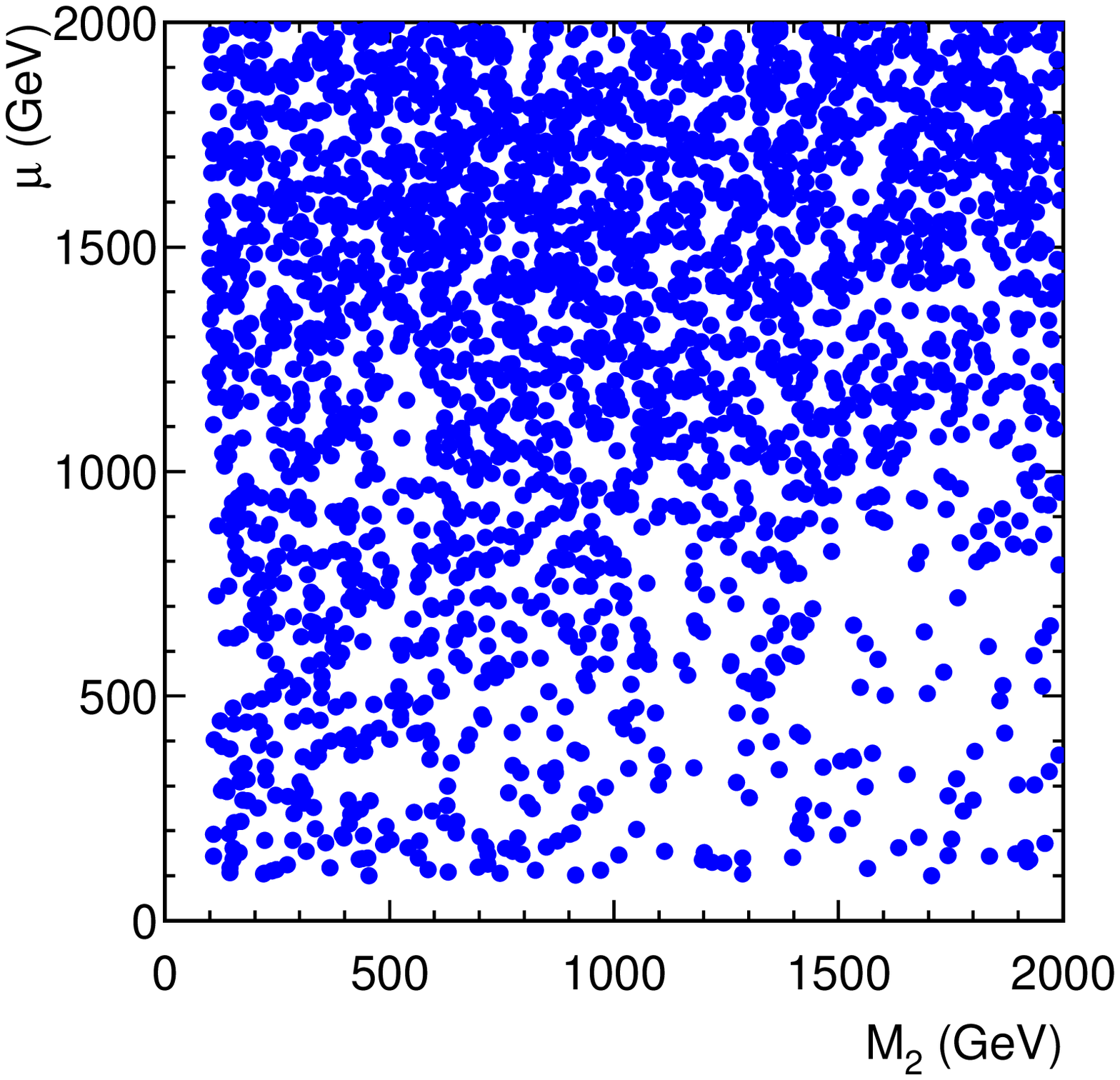}}
%\hfill
\subfloat[\label{dd:b}]{
\includegraphics[scale=1,width=7.5cm]{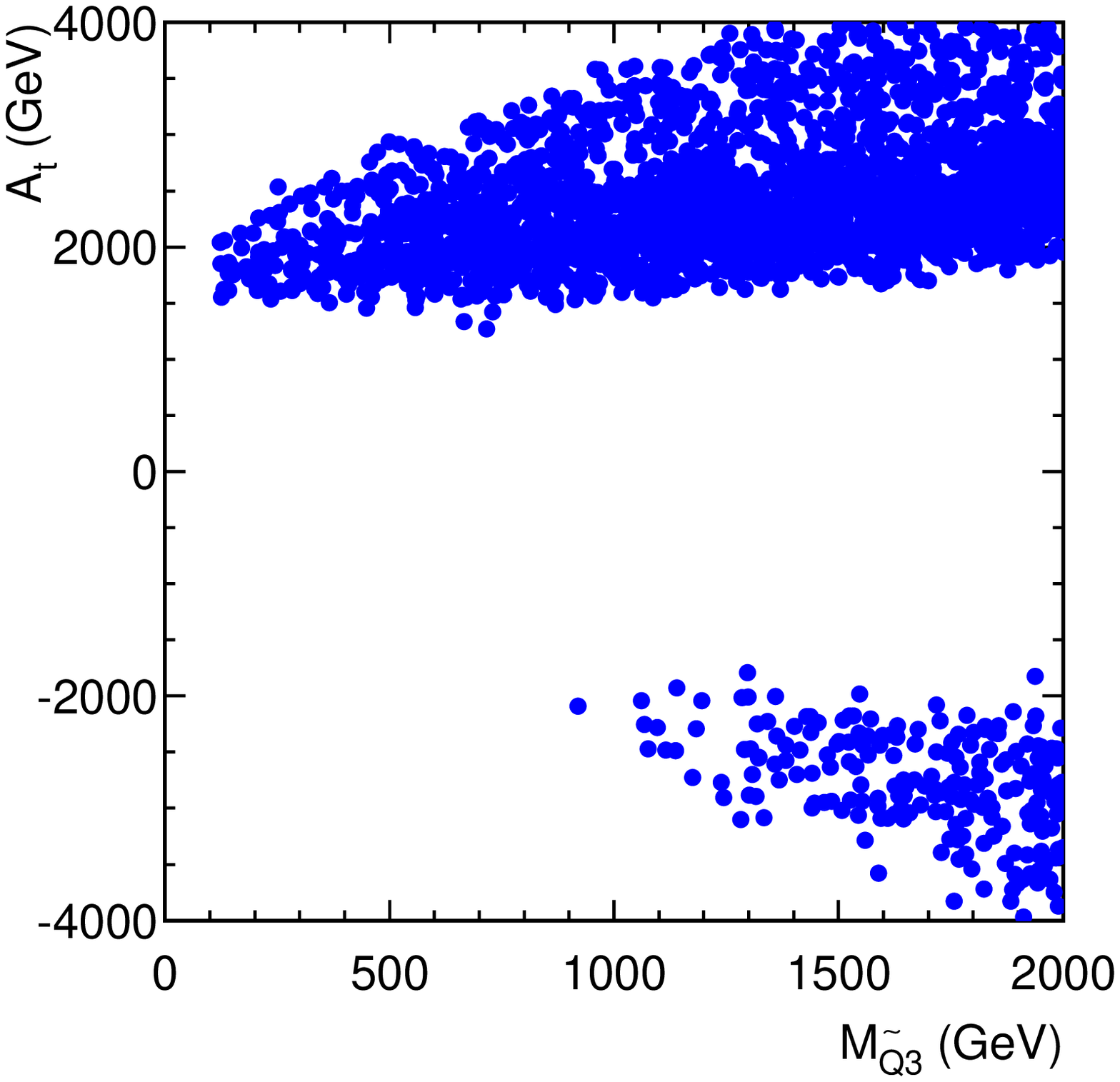}}
\\
\subfloat[\label{dd:c}]{
\includegraphics[scale=1,width=7.5cm]{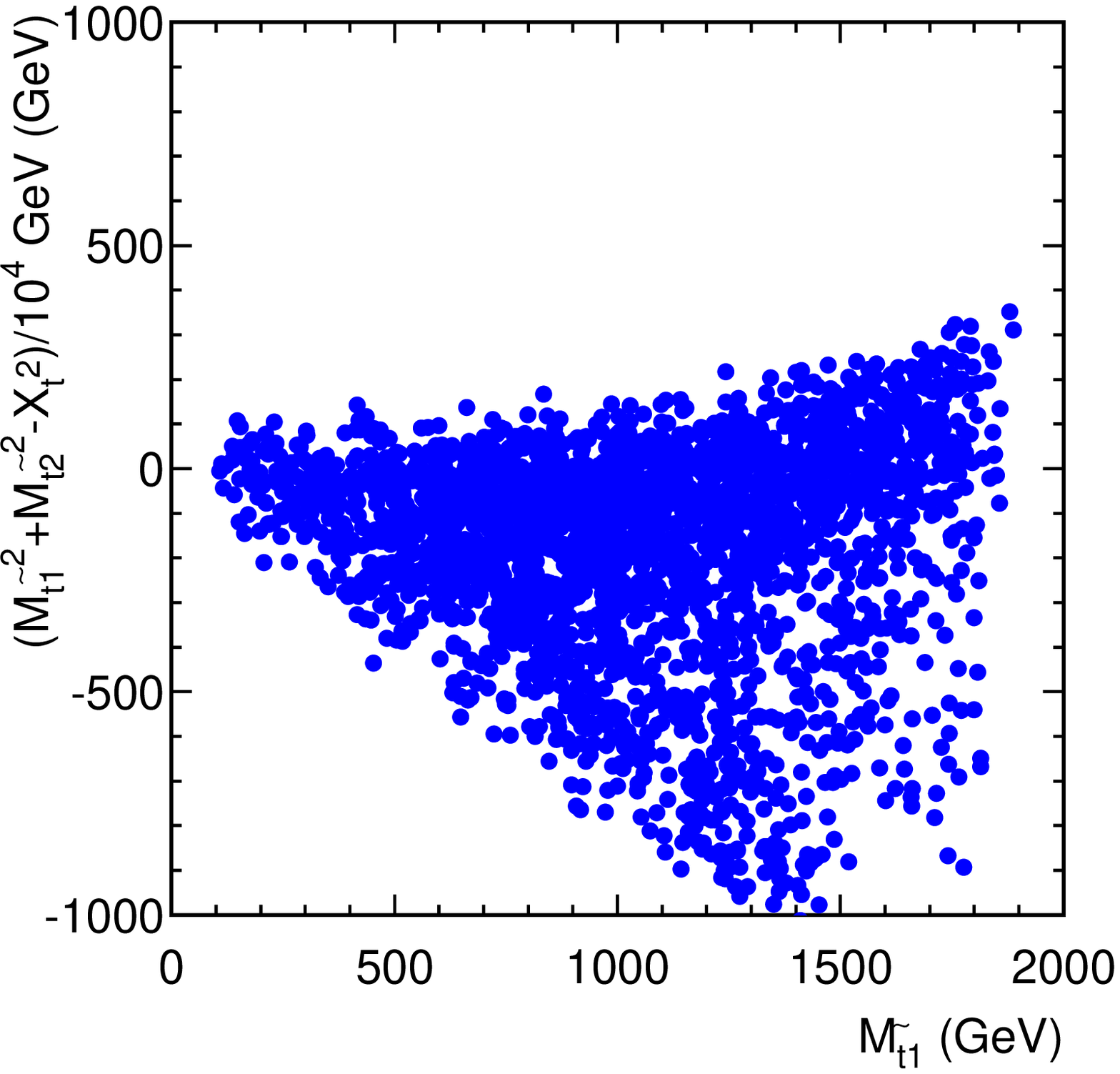}}
%\hfill
\subfloat[\label{dd:d}]{
\includegraphics[scale=1,width=7.5cm]{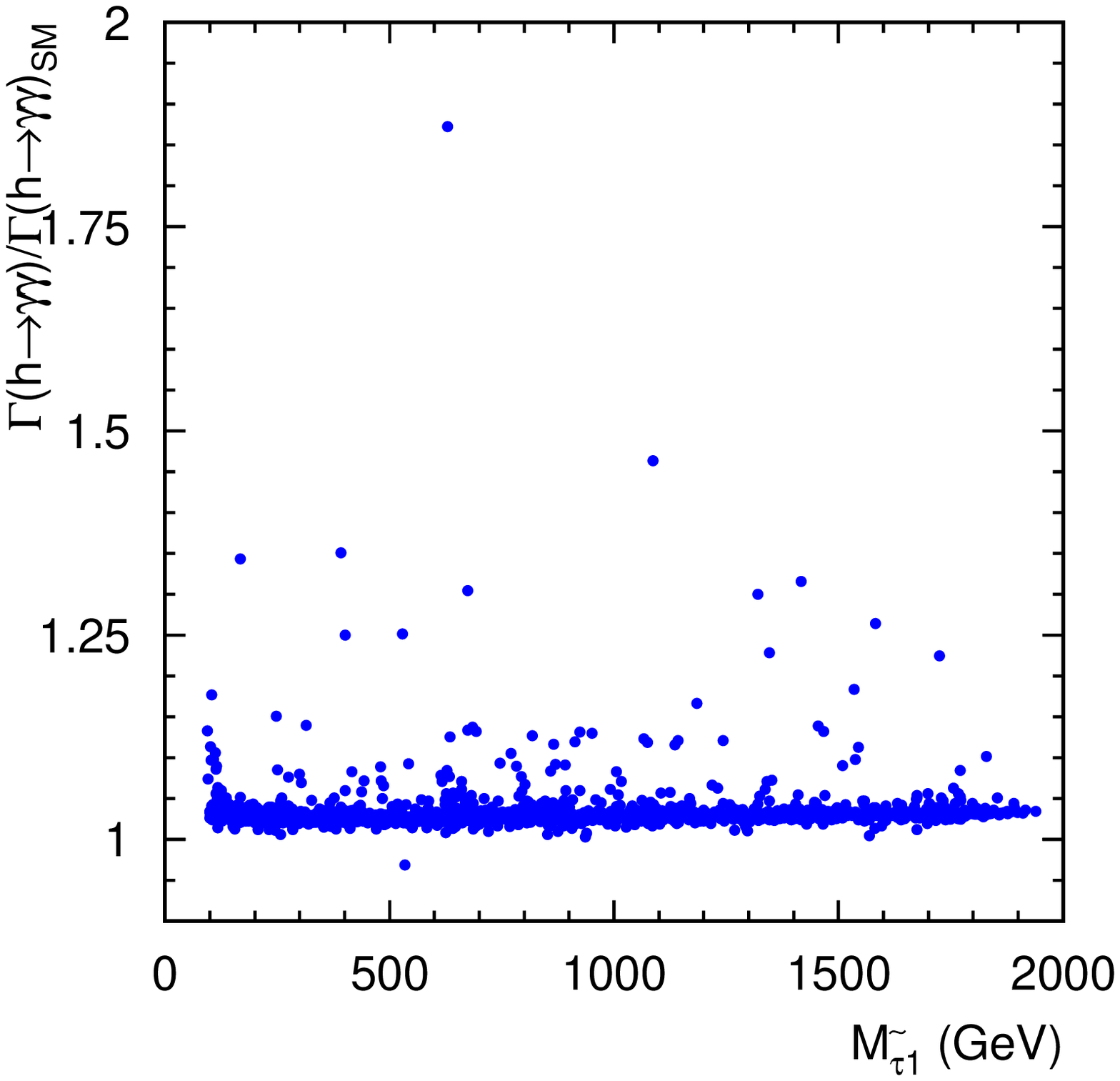}}
\end{center}
\caption{(a) $\mu$ versus $M_2$, (b) $A_t$ versus $m_{\tilde{Q}_3}$, (c) $(m_{\tilde{t}_1}^2+m_{\tilde{t}_2}^2-X_t^2)/10^4 \ {\rm GeV}$ versus $m_{\tilde{t}_1}$ and (d) $\Gamma(h^0\to \gamma\gamma)/\Gamma(h^0\to \gamma\gamma)_{SM}$ versus $m_{\tilde{\tau}_1}$.} \label{param}
\end{figure}

As well discussed before, there are two main mechanisms leading to a simultaneous enhancement of the diphoton production rate in the MSSM~\cite{Heinemeyer}.
%Firstly, a reduction of the $h^0\to b\bar{b}$ partial width is effectively a reduction of the total decay width as the largest contribution, which leads to enhance the $\gamma\gamma$ rate.
Firstly, the largest partial contribution to the total width of SM-like Higgs decay, namely $\Gamma(h^0\to b\bar{b})$, would decrease if the bottom Yukawa is enhanced. As a result, the total decay width of $h^0$ will be reduced and thus the $\gamma\gamma$ rate gets enhancement.
Figs.~\ref{param} (a) and (b) show the allowed parameter space relevant for the SM-like Higgs production: (a) $\mu$ versus $M_2$ and (b) $A_t$ versus $m_{\tilde{Q}_3}$. The current Higgs bounds strongly favor relatively large $\mu$ and positive $A_t$ with $|A_t| \gtrsim 2$ TeV. This is because large positive product $\mu A_t$ leads to a large positive radiative correction to bottom Yukawa which is needed to suppress $\Gamma(h^0\to b\bar{b})$ so as to enhance $\sigma(gg\to h^0\to \gamma\gamma)$~\cite{Christensen:2012ei,Heinemeyer}.
%There are still narrow region corresponding to low $\mu$ and negative $A_t$, where it is known that light stau plays
%an important role for the enhancement of $\Gamma(h^0\to \gamma\gamma)$~\cite{stau}.

%, or small $M_2$ and $\mu$.

The second mechanism is due to the effect of SUSY particles in the direct enhancement of the $\Gamma(h^0\to gg/\gamma\gamma)$, for instance light stop and stau~\cite{carlosstau}.
The stop loop contributions to the $gg$ and $\gamma\gamma$ amplitudes are approximately proportional to~\cite{stopamp,carlosstopstau}
\begin{eqnarray}
%\delta A_{gg,\gamma\gamma}\propto \pm {m_t^2\over m_{\tilde{t}_1}^2m_{\tilde{t}_2}^2}\left(m_{\tilde{t}_1}^2+m_{\tilde{t}_2}^2-X_t^2\right).
\pm {m_t^2\over m_{\tilde{t}_1}^2m_{\tilde{t}_2}^2}\left(m_{\tilde{t}_1}^2+m_{\tilde{t}_2}^2-X_t^2\right).
\label{stophiggsprod}
\end{eqnarray}
%Hence, for values of the mixing parameter $X_t^2>(<) \ m_{\tilde{t}_1}^2+m_{\tilde{t}_2}^2$, the stops lead to a reduction (enhancement) of the gluon fusion
%Higgs production and an enhancement (reduction) of the Higgs to diphoton decay width.
Hence, we show the stop effect in Higgs production described in Eq.~(\ref{stophiggsprod}) in Fig.~\ref{param} (c) in the plane of $(m_{\tilde{t}_1}^2+m_{\tilde{t}_2}^2-X_t^2)/10^4 \ {\rm GeV}$ versus $m_{\tilde{t}_1}$.
For light stop, as one can see, the enhanced contribution of stop in the $\Gamma(h^0\to \gamma\gamma)$ dominates over the reduction in the gluon fusion production such that for $gg\to h^0\to \gamma\gamma$ rate being above 0.8 of the SM value. Moreover, an enhancement of $\Gamma(h^0\to \gamma\gamma)/\Gamma(h^0\to \gamma\gamma)_{SM}$ as large as a factor of 1.25 is possible as a result of light stau effect in the loop, as seen in Fig.~\ref{param} (d).

%\begin{figure}[tb]
%\begin{center}
%\minigraph{5cm}{-0.15in}{(a)}{../plots/m2mu.eps}}
%\hfill
%\minigraph{5cm}{-0.15in}{(b)}{../plots/Atmsq3.eps}}\\
%\minigraph{5cm}{-0.15in}{(c)}{../plots/Atma.eps}}
%\minigraph{5cm}{-0.15in}{(d)}{../plots/Atama.eps}}
%\includegraphics[scale=1,width=8cm]{plot/ttb_MC.eps}
%\end{center}
%\caption{$\tan\beta$ vs. $m_A$. Red: with SM Higgs mass and Higgs
%search constraints, Green: with additional B rare decay
%constraints.} \label{a}
%\end{figure}

%\begin{figure}[tb]
%\begin{center}
%\includegraphics[scale=1,width=8cm]{../plots/m2mu.eps}
%\includegraphics[scale=1,width=8cm]{../plots/msq3-mstr.eps}
%\end{center}
%\caption{$M_2$ vs. $\mu$ and $M_{\tilde{t}_R}$ vs. $M_{\tilde{Q}_3}$.} \label{a}
%\end{figure}

%%%%%%%%%%%%%%%%%%%%%%%%%%%%%%%%%%%%
%\subsection{SUSY Spectrum}

\subsection{Discussion of SUSY Sparticle Searches}

Additional constraints come from direct sparticle searches, for instance stop and sbottom.
In principle, the stop and sbottom mass limit drops lower for small mass difference between
the stop/sbottom and the Bino LSP. One can always tune the free Bino mass to be large enough to give soft decay products and thus evade the
stop/sbottom search limits. Recently, ATLAS reported that light stops with $m_{\tilde{t}_1}\lesssim 200$ GeV and any kinematically allowed neutralino LSP mass are essentially excluded if $BR(\tilde{t}_1\to c\tilde{\chi}_1^0)=100\%$~\cite{atlasstop-c}. However, this bound could be weakened if other decay mode with lighter sparticle, such as $\tilde{t}_1\to \tilde{\tau}_1\nu_\tau b$, overwhelms $\tilde{t}_1\to c\tilde{\chi}_1^0$ as pointed out in Ref.~\cite{carlosstopstau}. Also, if Bino mass is not that large and $m_{\tilde{t}_1}-m_{\tilde{\chi}_1^0}>m_W+m_b$ ($m_t$), the main decay mode is given by $\tilde{t}_1\to bW^+ \tilde{\chi}_1^0 \ (t\tilde{\chi}_1^0)$. We then have freedom for Bino mass to survive light stop, given the gap between stop bound and kinematic limit.

ATLAS also released that any sbottom with mass less than 650 GeV is not allowed if $m_{\tilde{\chi}_1^0}<100$ GeV and $BR(\tilde{b}_1\to b\tilde{\chi}_1^0)=100\%$~\cite{atlassbottom}.
%In the small $m_{\tilde{Q}_3}$ case, light left-handed sbottom will appear in the spectrum. The $m_{\tilde{t}_1}$ and $m_{\tilde{b}_1}$ are then strongly correlated as $m_{\tilde{b}_1}\sim m_{\tilde{Q}_3}$ and the light stop mass is shifted by a fraction (about 30\%) of the top mass as shown in Eq.~(\ref{stopmass}). As a result, $m_{\tilde{b}_1}-m_{\tilde{\chi}_1^0}$ would be at least 60 GeV such that light sbottoms tend to be in conflict with the above limit.
For small values of $m_{\tilde{Q}_3}$, we have light left-handed sbottom in the spectrum as $m_{\tilde{b}_1}\sim m_{\tilde{Q}_3}$. Thus, this case tends to be in conflict with the above limit if $m_{\tilde{b}_1}-m_{\tilde{\chi}_1^0}\gtrsim 20$ GeV or $m_{\tilde{\chi}_1^0}<100$ GeV.
However, if Wino neutralino stays between sbottom and Bino LSP, the left-handed sbottom prefers to decay to it with $BR(\tilde{b}_1\to b\tilde{\chi}_2^0)$ being typically around 80\%-90\%~\cite{nondecoupling}, even though relatively suppressed by the available phase space. With the further decay of $\tilde{\chi}_2^0$ into $h^{0(\ast)}\tilde{\chi}_1^0$ or $Z^{(\ast)}\tilde{\chi}_1^0$, these longer decay chains give soft decay products and small missing energy undetected in the detector.
As a result, the current sbottom search would not highly restrict the small $m_{\tilde{Q}_3}$ case.
%Therefore, they will not cover the low mass scenario.

In addition, CMS put the lower limit on the $m_{\tilde{\chi}_1^\pm, \tilde{\chi}_2^0}$ to 330 GeV under the assumption of $m_{\tilde{\chi}_2^0}-m_{\tilde{\chi}_1^0}>m_Z$ and $BR(\tilde\chi_2^0 \rightarrow Z \tilde\chi_1^0)=BR(\tilde\chi_1^\pm \rightarrow W^\pm \tilde\chi_1^0)=100$\%~\cite{cmsgaugino}. This limit would not directly constrain the spectrum with small mass difference $m_{\tilde{\chi}_2^0}-m_{\tilde{\chi}_1^0}$ as well as possible suppression of chargino/neutralino decays.
%${\rm BR}(\tilde\chi_2^0 \rightarrow Z \tilde\chi_1^0)$.

%\begin{figure}[tb]
%\begin{center}
%\includegraphics[scale=1,width=8cm]{../plots/msb.eps}
%\includegraphics[scale=1,width=8cm]{../plots/mstlimit.eps}\\
%\includegraphics[scale=1,width=8cm]{../plots/mstanu.eps}
%\end{center}
%\caption{sparticle spectrum.} \label{a}
%\end{figure}

%%%%%%%%%%%%%%%%%%%%%%%%%%%%%%%%%%%%%%%%%%%%%%%%%%%%%%%%%%%%
\section{Heavy Higgs Decay and Search Sensitivity}
%%%%%%%%%%%%%%%%%%%%%%%%%%%%%%%%%%%%%%%%%%%%%%%%%%%%%%%%%%%%%%%

%%%%%%%%%%%%%%%%%%%%%%%%%%%%%%%%%%%%
%\subsection{SM-like Higgs Production}

%\begin{figure}[tb]
%\begin{center}
%\includegraphics[scale=1,width=8cm]{../plots/Wgaga-msta.eps}
%\includegraphics[scale=1,width=8cm]{../plots/Wgg-mst.eps}\\
%\includegraphics[scale=1,width=8cm]{../plots/WBR-msta.eps}
%\includegraphics[scale=1,width=8cm]{../plots/WBR-mst.eps}
%\end{center}
%\caption{Higgs production via gluon fusion and decay into $\gamma\gamma$ as a function of $m_{\tilde{t}_1}$ and $m_{\tilde{\tau}_1}$.} \label{a}
%\end{figure}

%%%%%%%%%%%%%%%%%%%%%%%%%%%%%%%%%%%
\subsection{Heavy Higgs Decay}

\begin{figure}[tb]
\begin{center}
\subfloat[\label{dd:a}]{
\includegraphics[scale=1,width=7.5cm]{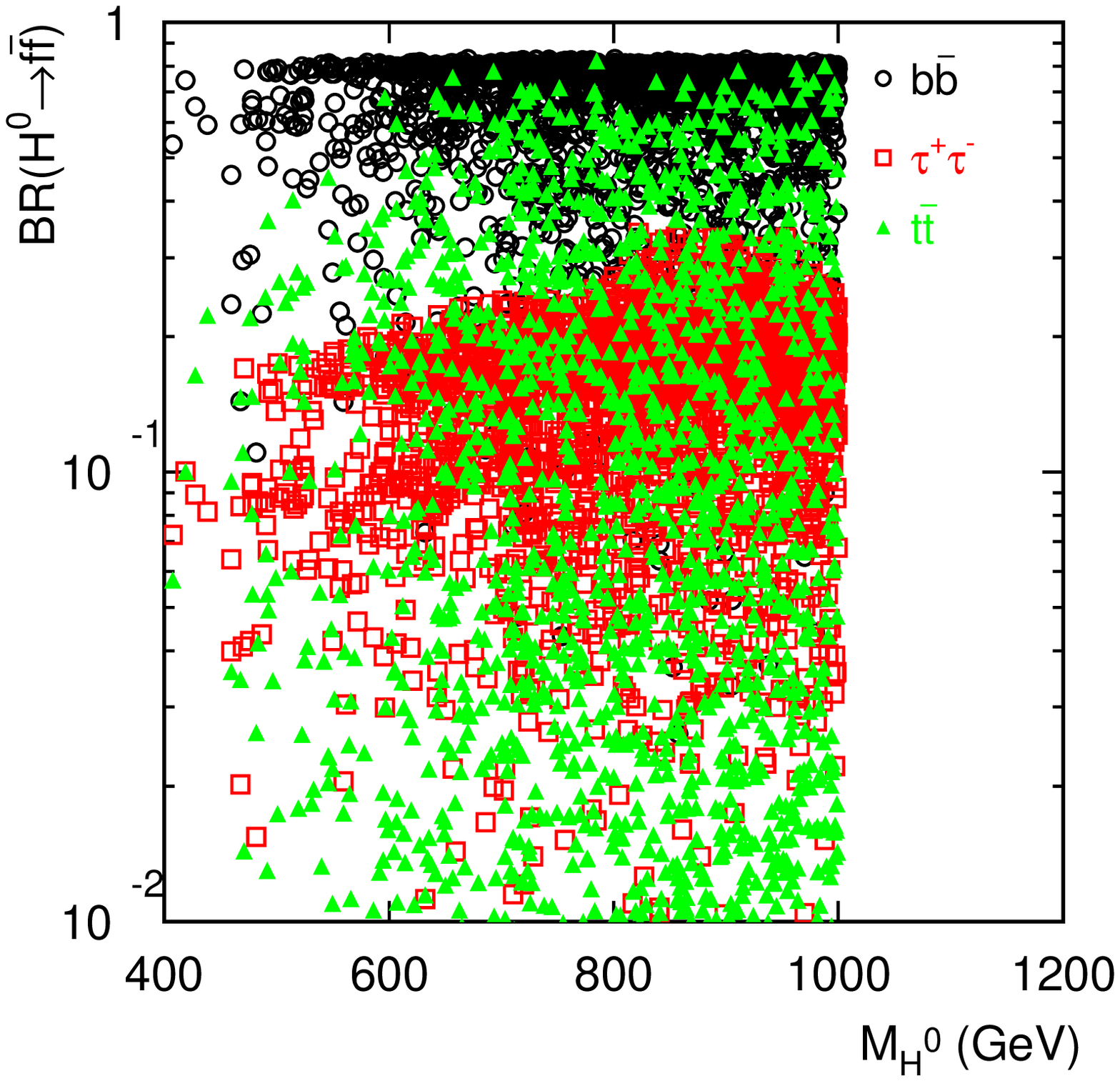}}
\subfloat[\label{dd:b}]{
\includegraphics[scale=1,width=7.5cm]{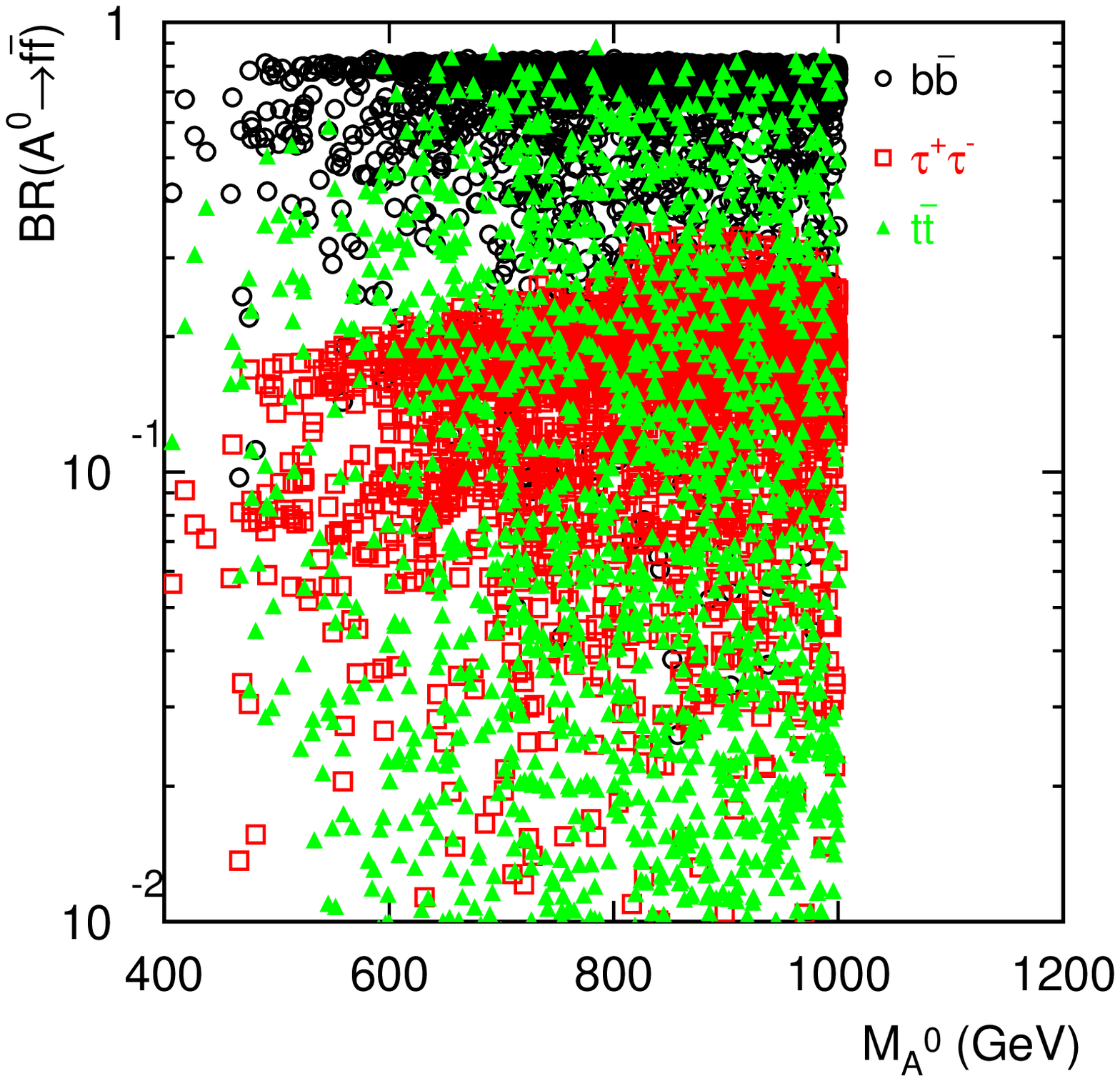}}
\end{center}
\caption{(a) $BR(H^0\to f\bar{f})$ vs. $m_{H^0}$ and (b) $BR(A^0\to f\bar{f})$ vs. $m_{A^0}$.} \label{ffa}
\end{figure}

\begin{figure}[tb]
\begin{center}
\subfloat[\label{dd:a}]{
\includegraphics[scale=1,width=8.5cm]{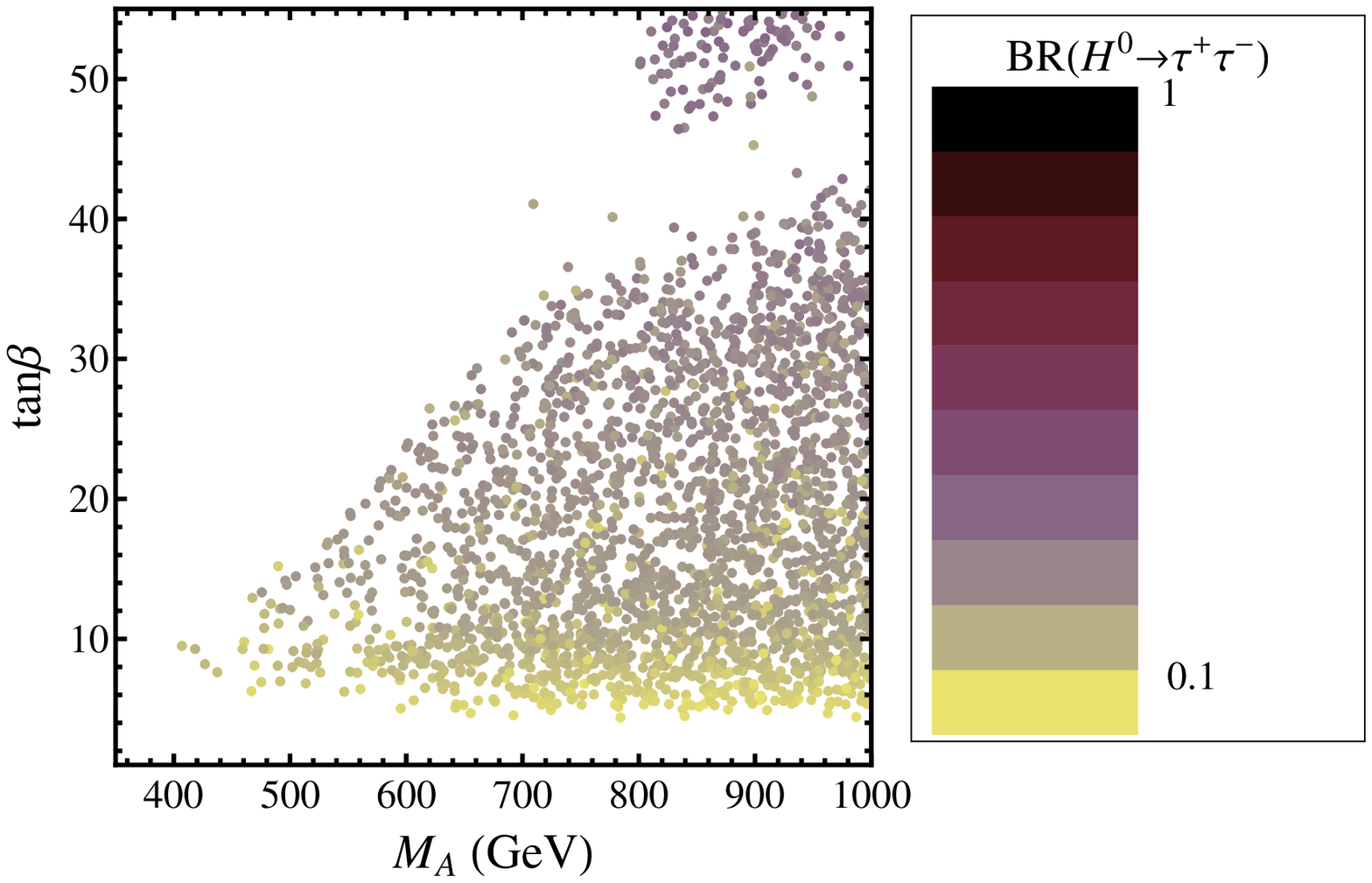}}
\subfloat[\label{dd:b}]{
\includegraphics[scale=1,width=8.5cm]{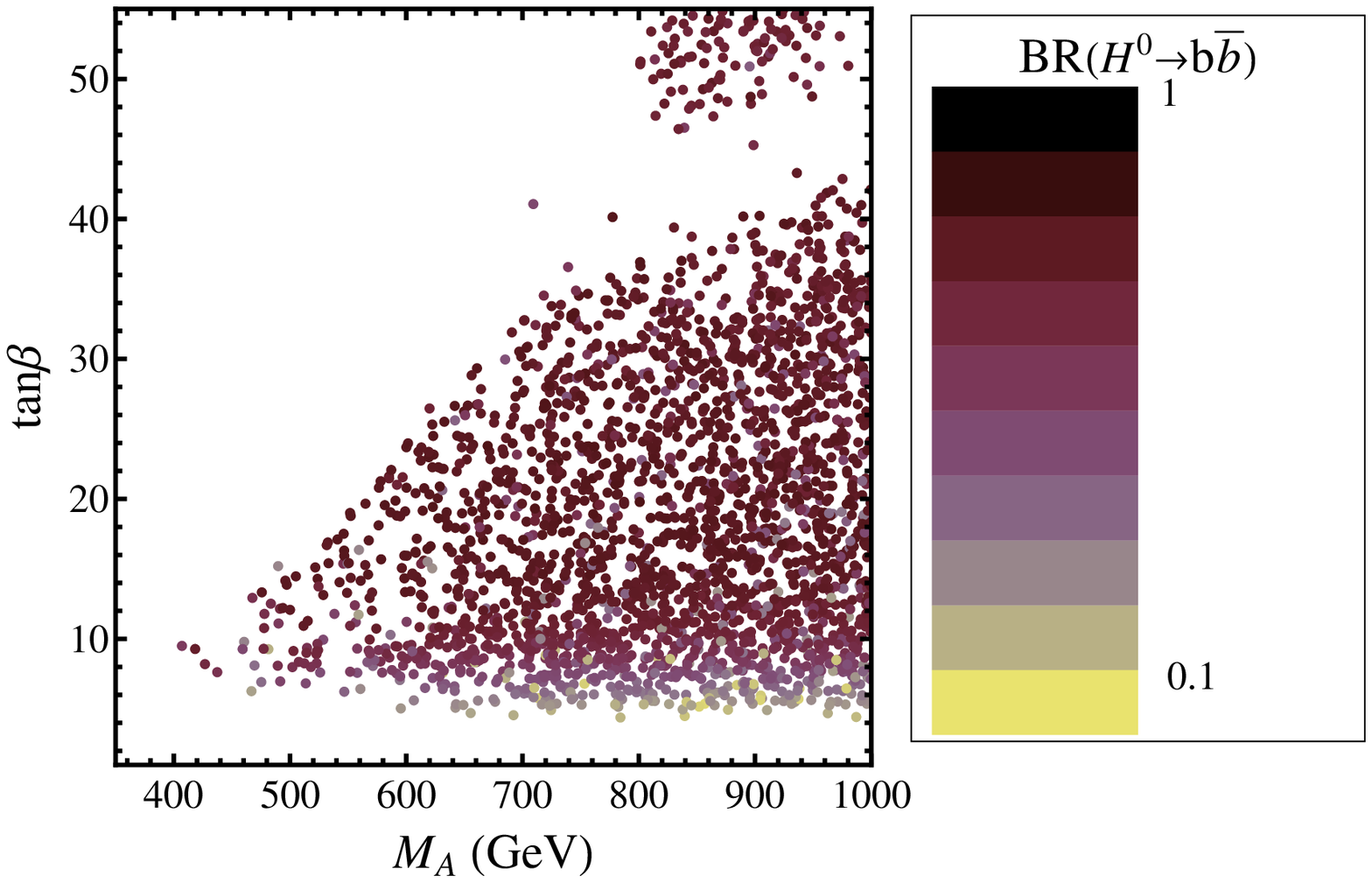}}\\
\subfloat[\label{dd:c}]{
\includegraphics[scale=1,width=8.5cm]{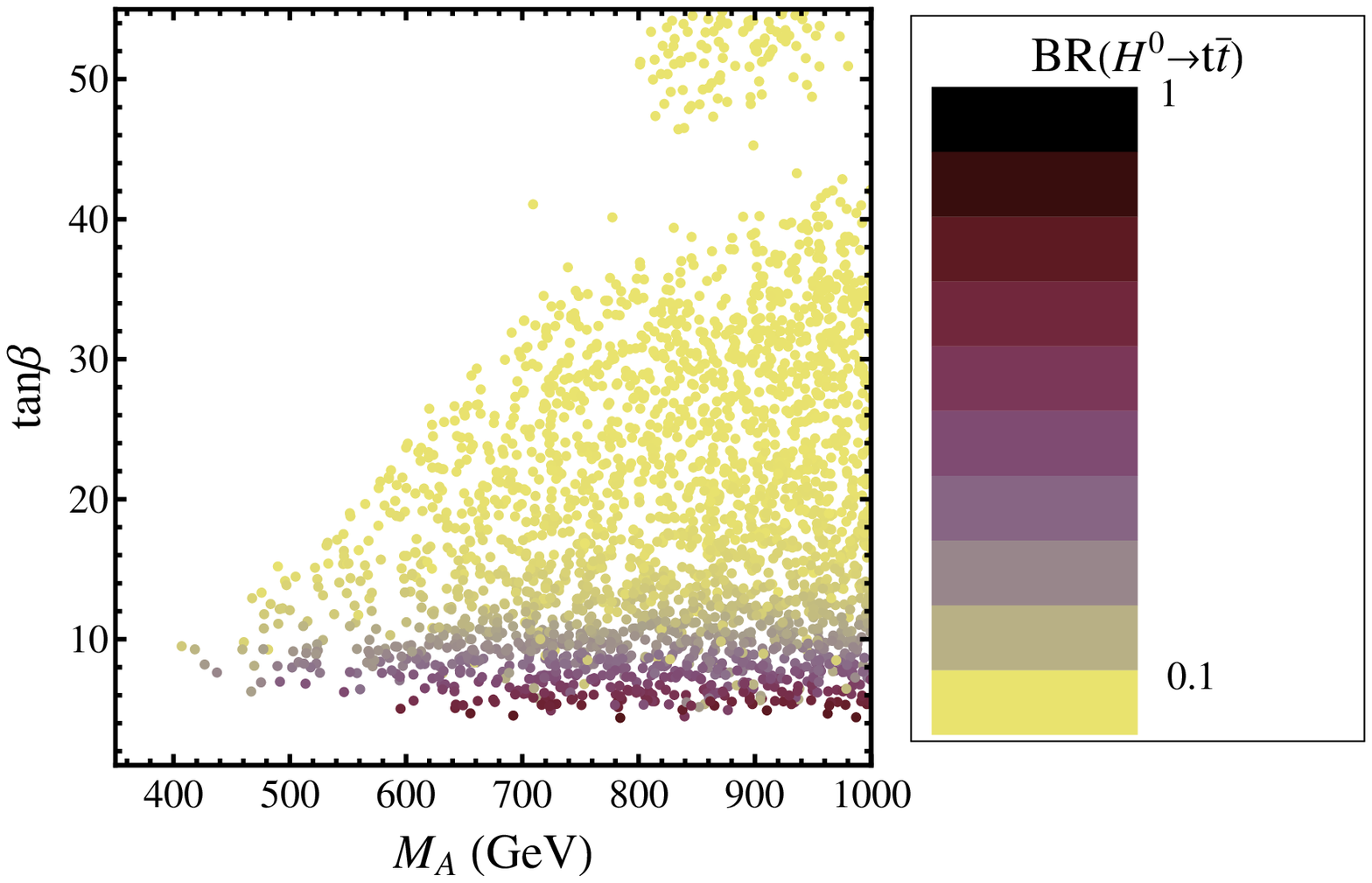}}
\end{center}
\caption{$BR(H^0\to f\bar{f})$ in the plane of $\tan\beta$ vs. $M_A$, for (a) $f=\tau$, (b) $f=b$ and (c) $f=t$. The color scale gives the branching fraction of $H^0\to f\bar{f}$ decay.} \label{ffb}
\end{figure}

In the decoupling limit, the heavy non SM-like Higgses $H^0$, $A^0$ and $H^\pm$ have rich decay modes, especially in the small $\tan\beta$ regime.
Figs.~\ref{ffa} and \ref{ffb} show the branching ratios of heavy neutral Higgs bosons decay into fermion pairs. In this limit, the $H^0/A^0$ coupling to the top quarks is suppressed by $1/\tan\beta$, while the couplings to bottom quarks and tau leptons are enhanced by $\tan\beta$. As seen in Fig.~\ref{ffa}, a majority of points have $BR(H^0/A^0\to b\bar{b})\sim 80\%$ and $BR(H^0/A^0\to \tau^+\tau^-)\sim 20\%$. However, for exceptional significant points in Figs.~\ref{ffa} and \ref{ffb}, the $H^0/A^0\to t\bar{t}$ mode could be dominant for $\tan\beta\lesssim 10$ in particular.

\begin{figure}[tb]
\begin{center}
\subfloat[\label{dd:a}]{
\includegraphics[scale=1,width=7.5cm]{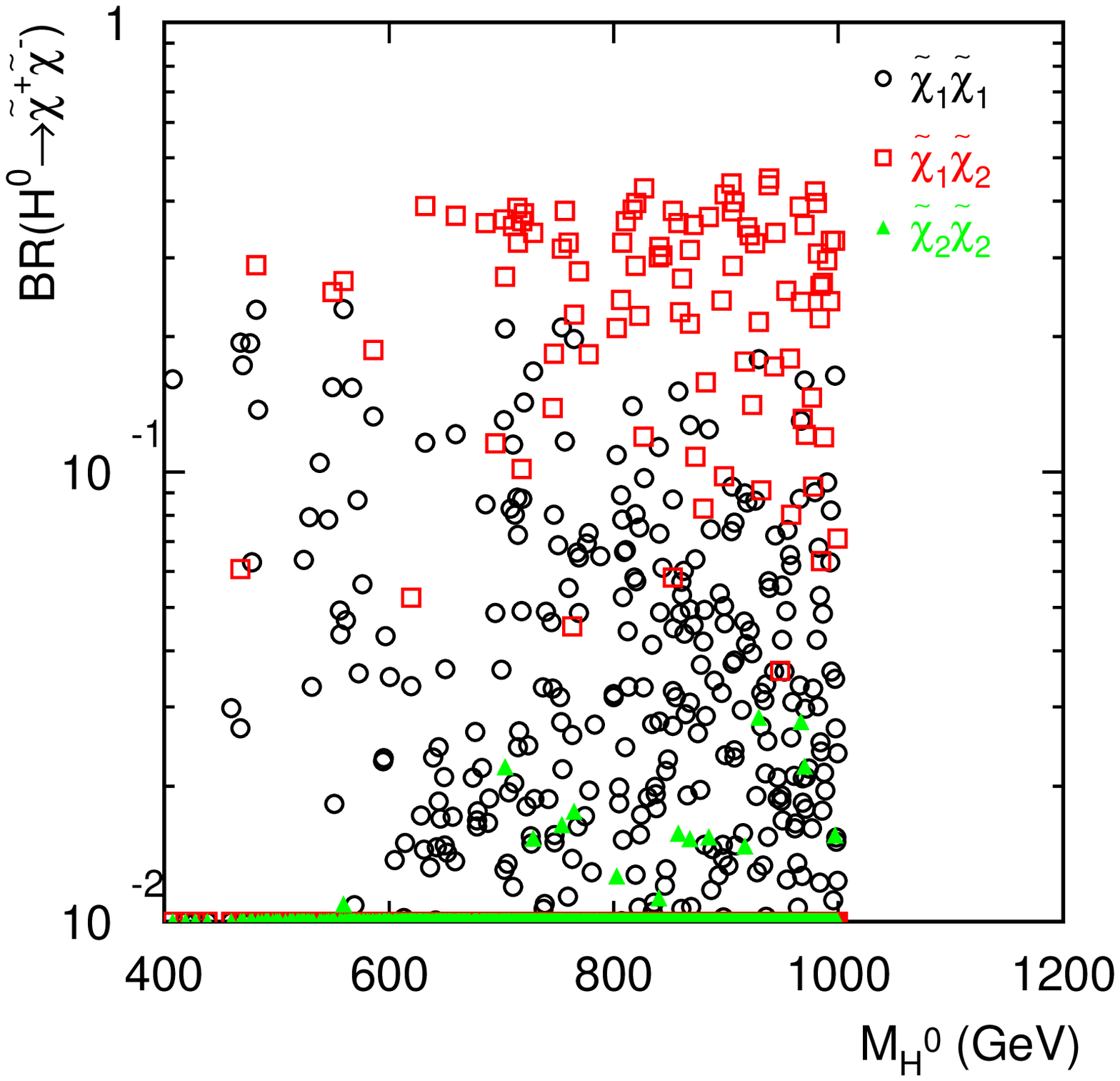}}
\subfloat[\label{dd:b}]{
\includegraphics[scale=1,width=7.5cm]{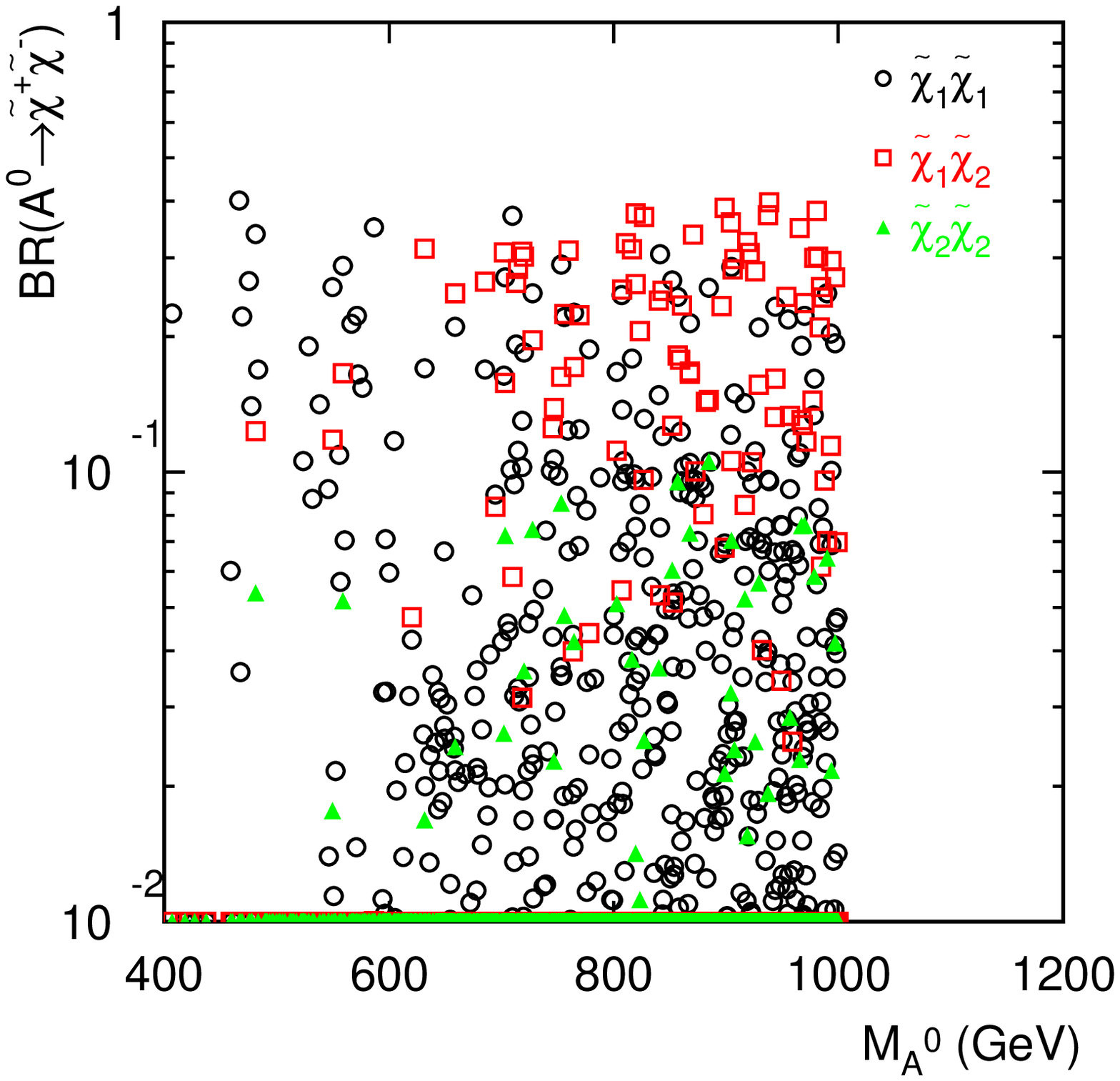}}
\end{center}
\caption{(a) $BR(H^0\to \tilde{\chi}^\pm_i\tilde{\chi}^\mp_j)$ vs.
$m_{H^0}$ and (b) $BR(A^0\to \tilde{\chi}^\pm_i\tilde{\chi}^\mp_j)$
vs. $m_{A^0}$.} \label{chpma}
\end{figure}

\begin{figure}[tb]
\begin{center}
\subfloat[\label{dd:a}]{
\includegraphics[scale=1,width=8.5cm]{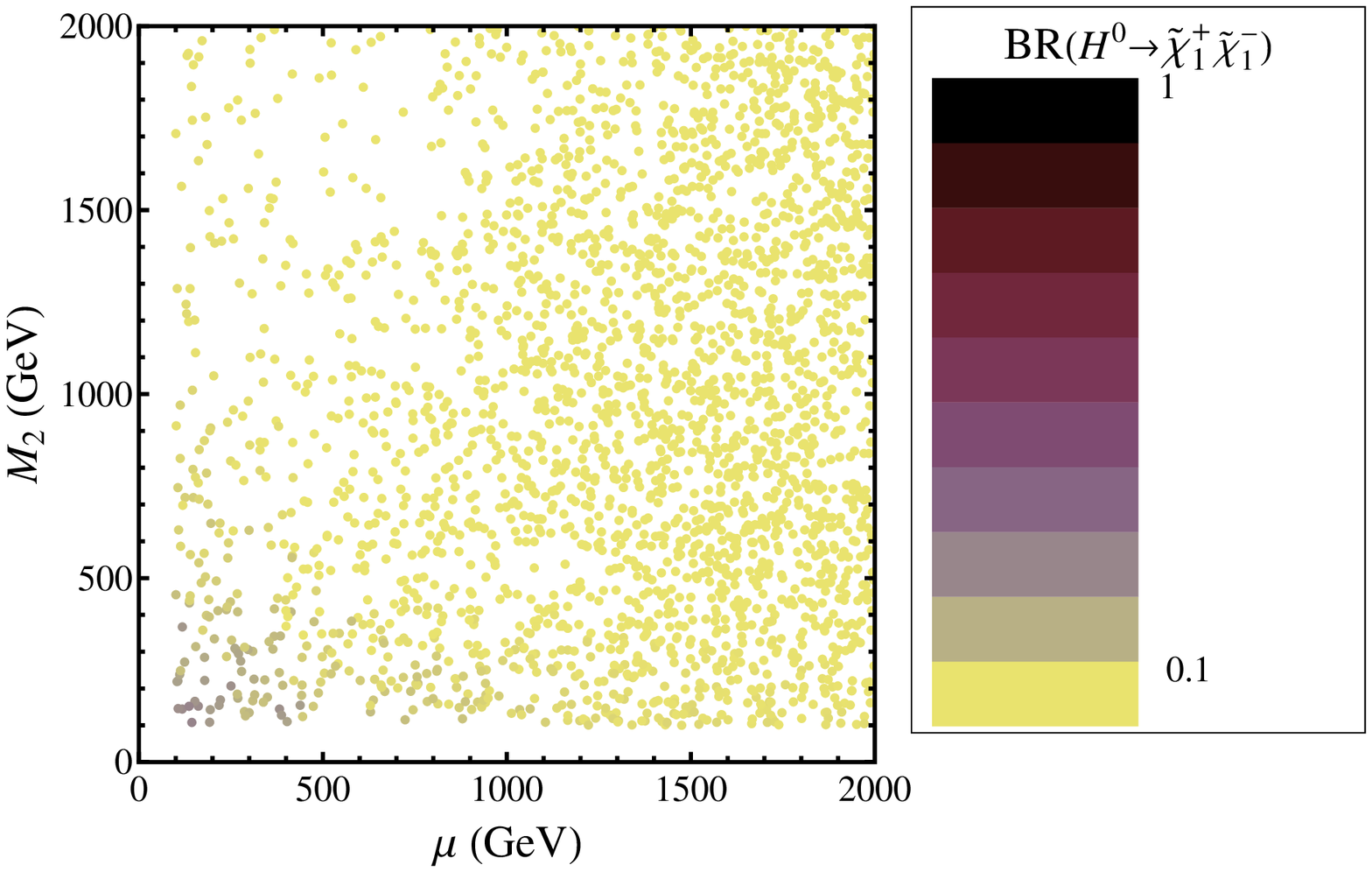}}
\subfloat[\label{dd:b}]{
\includegraphics[scale=1,width=8.5cm]{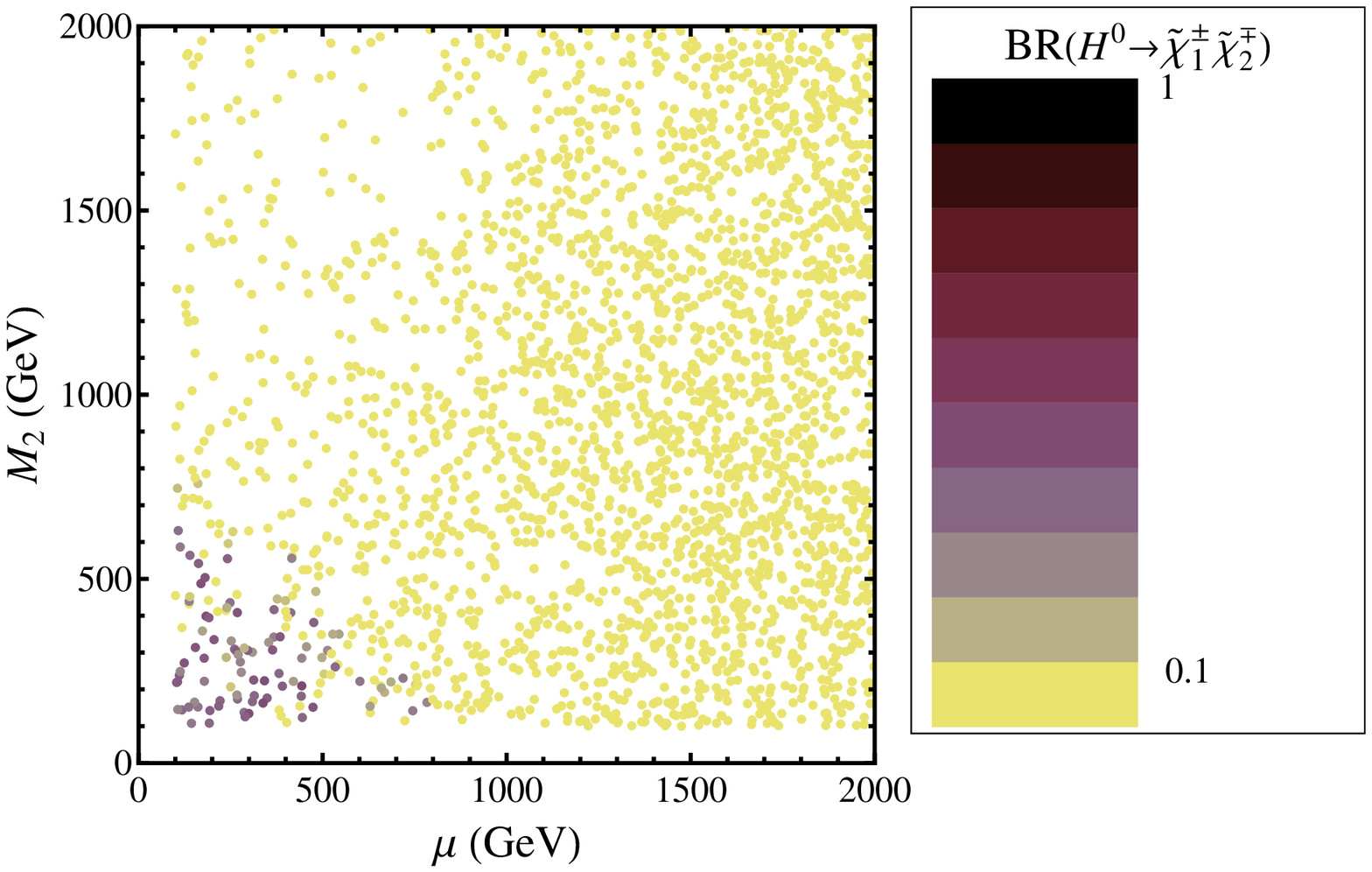}}
\end{center}
\caption{$BR(H^0\to \tilde{\chi}^\pm_i\tilde{\chi}^\mp_j)$ in the plane of $M_2$ vs. $\mu$, for (a) $i=j=1$ and (b) $i=1,j=2$. The color scale gives the branching fraction of $H^0\to \tilde{\chi}^\pm_i\tilde{\chi}^\mp_j$ decay.} \label{chpmb}
\end{figure}

\begin{figure}[tb]
\begin{center}
\subfloat[\label{dd:a}]{
\includegraphics[scale=1,width=7.5cm]{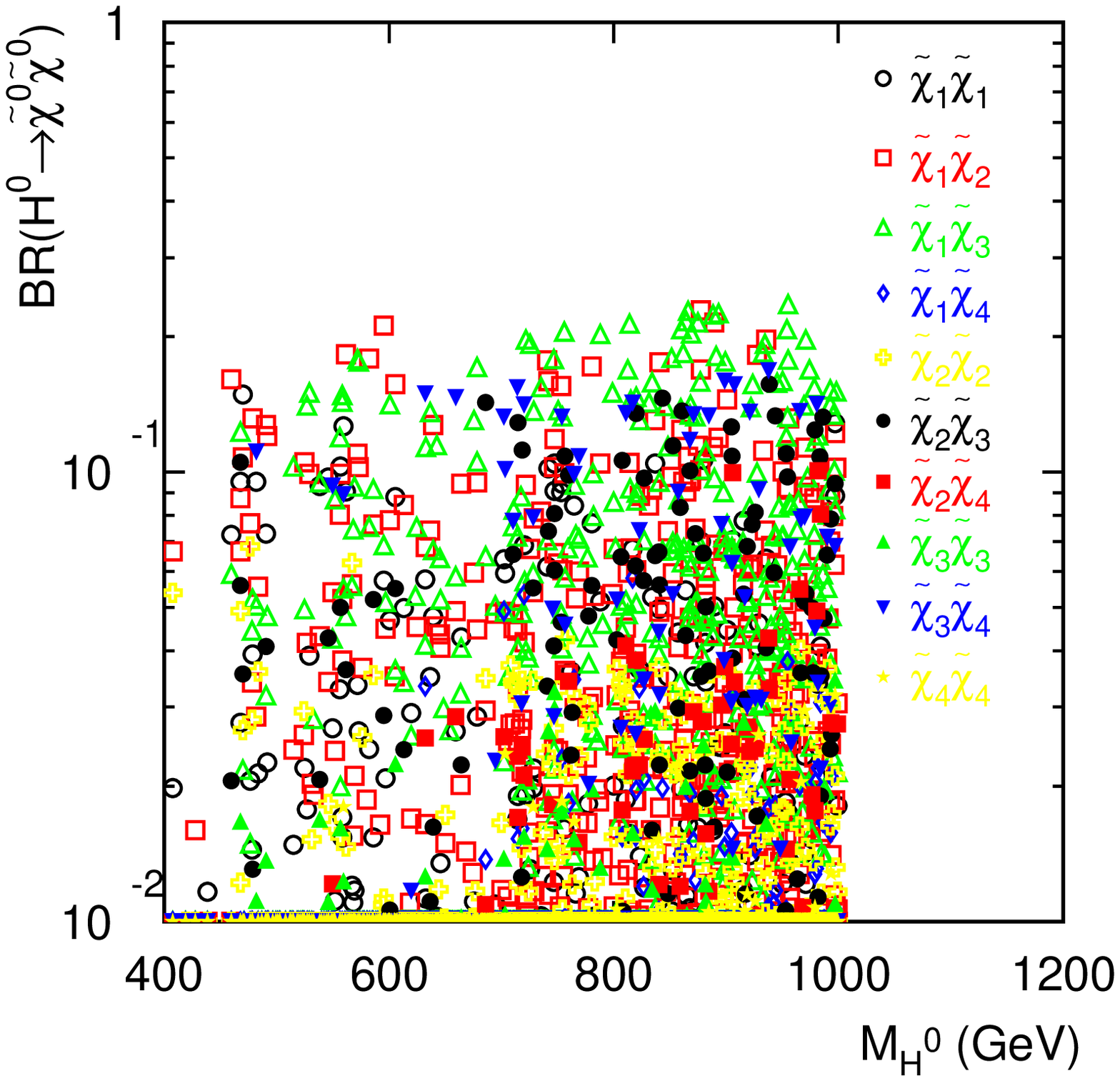}}
\subfloat[\label{dd:b}]{
\includegraphics[scale=1,width=7.5cm]{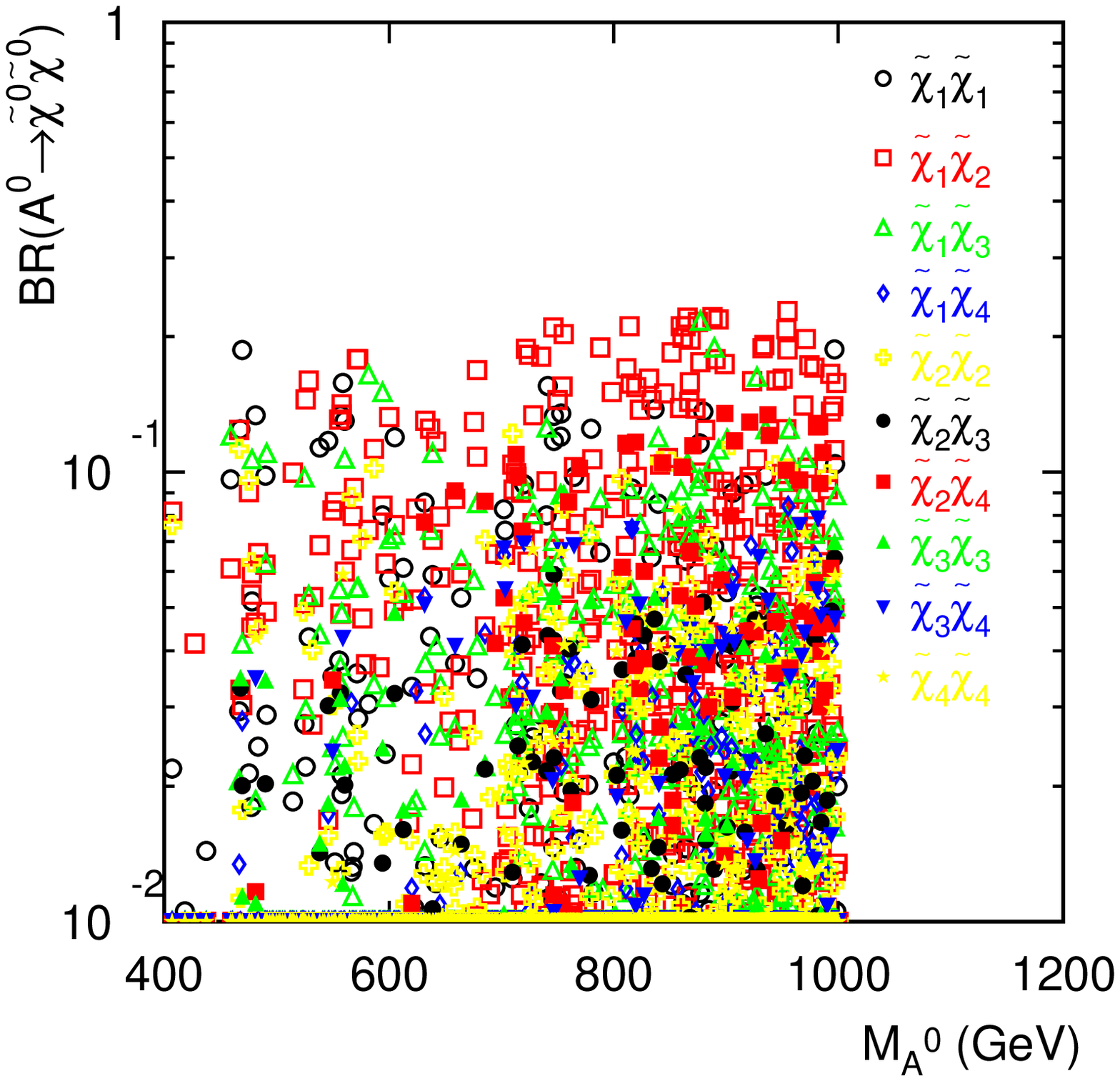}}
\end{center}
\caption{(a) $BR(H^0\to \tilde{\chi}^0_i\tilde{\chi}^0_j)$ vs.
$m_{H^0}$ and (b) $BR(A^0\to \tilde{\chi}^0_i\tilde{\chi}^0_j)$ vs.
$m_{A^0}$.} \label{ch0a}
\end{figure}

\begin{figure}[tb]
\begin{center}
\subfloat[\label{dd:a}]{
\includegraphics[scale=1,width=7.5cm]{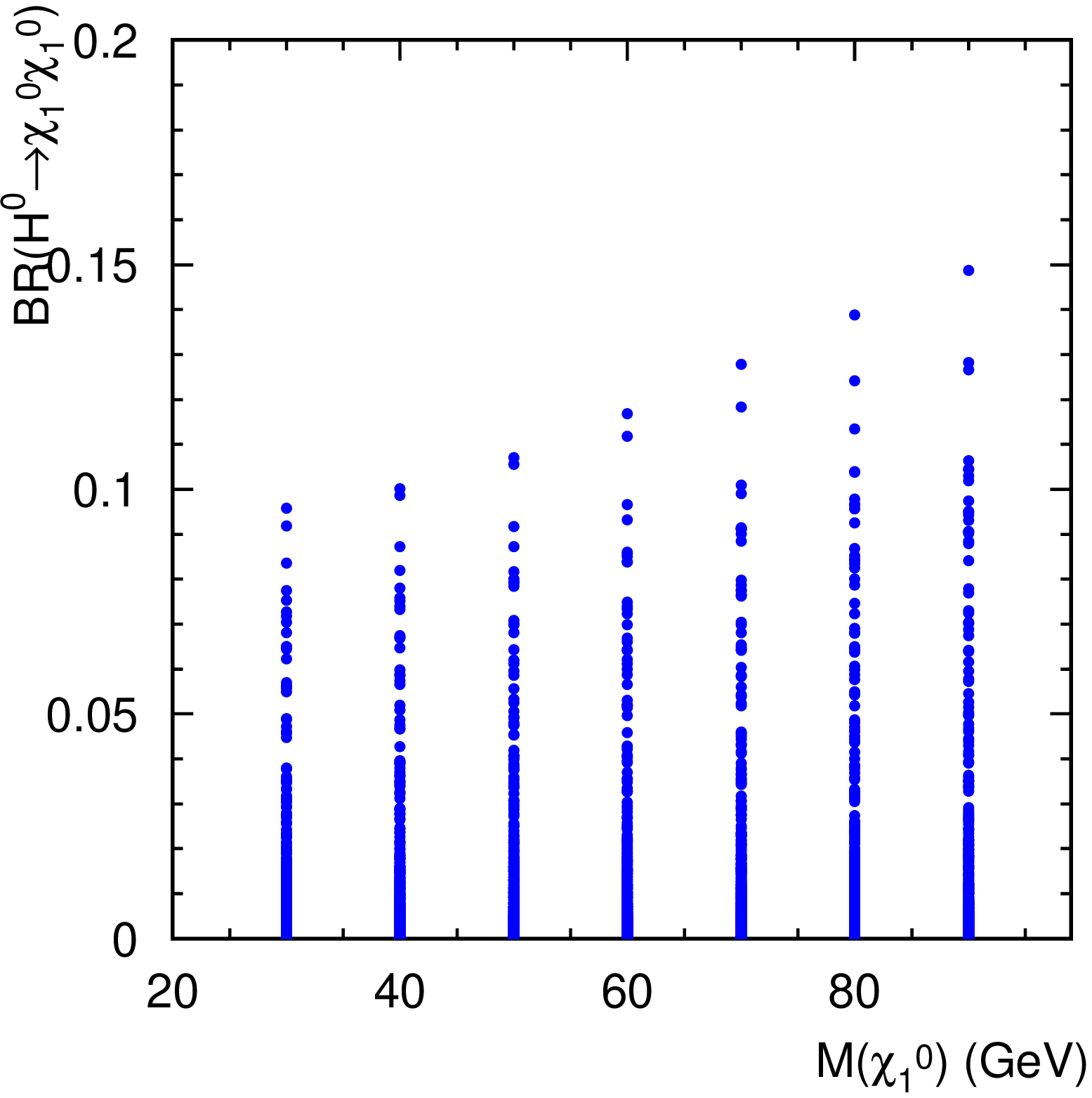}}
\subfloat[\label{dd:b}]{
\includegraphics[scale=1,width=7.5cm]{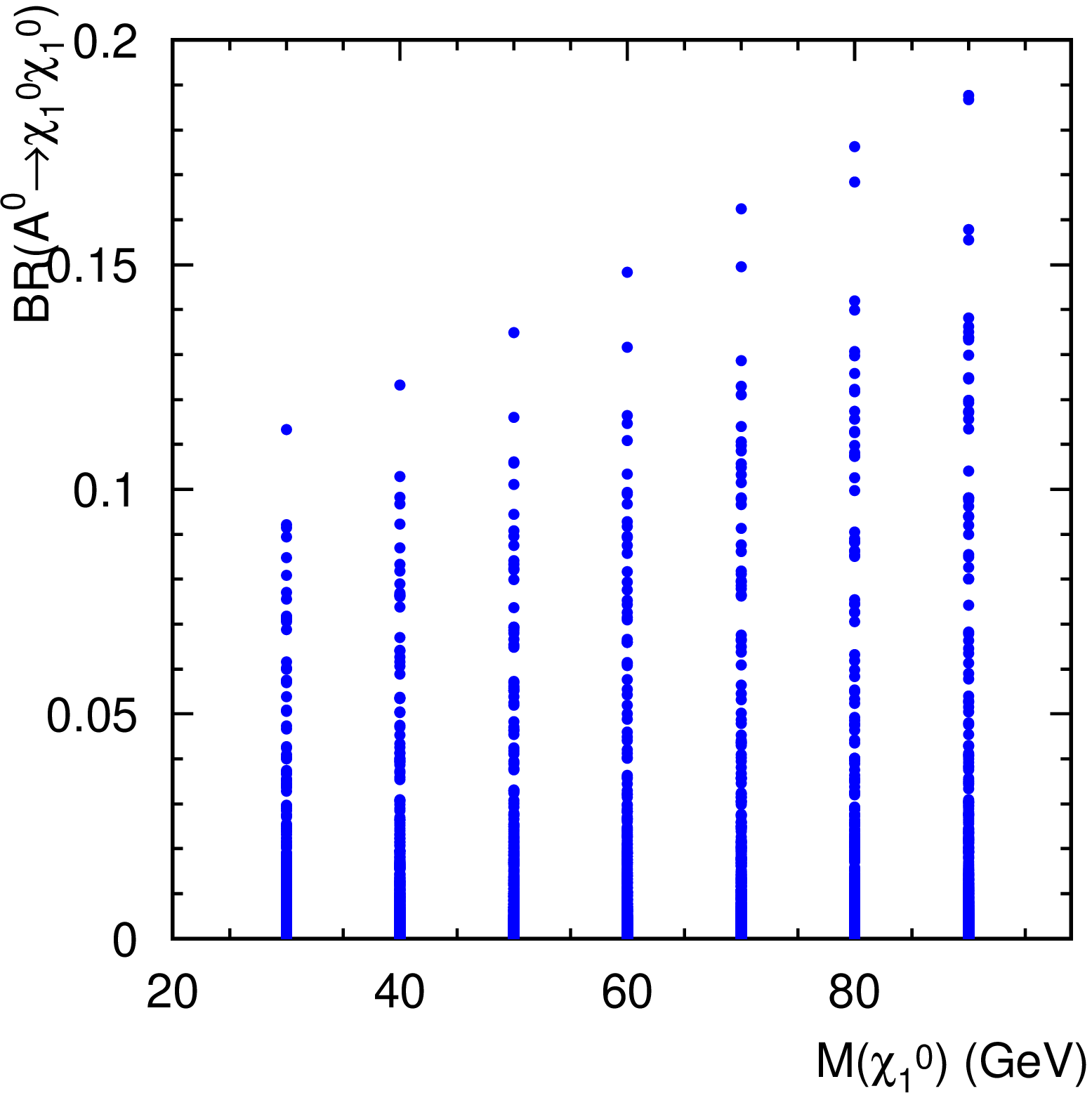}}
\end{center}
\caption{(a) $BR(H^0\to \tilde{\chi}^0_1\tilde{\chi}^0_1)$ vs. $m_{\tilde{\chi}_1^0}$ and (b) $BR(A^0\to \tilde{\chi}^0_1\tilde{\chi}^0_1)$ vs.
$m_{\tilde{\chi}_1^0}$.} \label{ch0b}
\end{figure}

Fig.~\ref{param} (a) shows that small values of $\mu, M_2$ are allowed. We thus expect kinematically occurred heavy Higgs decay into pairs of chargino and neutralino. The MSSM Higgs bosons mainly couple to mixtures of higgsino and gaugino components~\cite{Djouadi:2005gj}. Therefore, for $\mu \gg M_{1,2}$ or $\mu \ll M_{1,2}$, the decays of the heavy Higgs bosons into pairs of pure gaugino or higgsino are strongly suppressed.
The mixed decay $H^0/A^0\to \tilde{\chi}^\pm_1\tilde{\chi}^\mp_2,\tilde{\chi}^0_{1,2}\tilde{\chi}^0_{3,4}$ will then have significant branching fractions. For $\mu\sim M_2$, on the other hand, all the heavy Higgses have comparable decay rates into chargino/neutralino.
%decay channels occur at comparable rates when they are kinematically allowed.
We show the $BR(H^0/A^0\to \tilde{\chi}^\pm_i\tilde{\chi}^\mp_j)$ and $BR(H^0/A^0\to \tilde{\chi}^0_i\tilde{\chi}^0_j)$ in Figs.~\ref{chpma}, \ref{chpmb} and \ref{ch0a}. One can see that the branching ratio of chargino and neutralino decay mode can reach at most 40\% and 20\%, respectively.
In particular, the invisible decay mode of heavy higges, namely $H^0/A^0\to \tilde{\chi}^0_1\tilde{\chi}^0_1$, relies on the arbitrary Bino LSP mass and has important implication for the dark matter candidate search at the LHC. We display $BR(H^0/A^0\to \tilde{\chi}^0_1\tilde{\chi}^0_1)$ as a function of $m_{\tilde{\chi}^0_1}$ in Fig.~\ref{ch0b}. The $BR(H^0(A^0)\to \tilde{\chi}^0_1\tilde{\chi}^0_1)$ increases with increasing Bino LSP mass and is between 10\%-15\% (20\%) for $30<m_{\tilde{\chi}^0_1}<100$ GeV. This invisible decay mode can be tested through mono-b jet signature in $gb\to bH^0/A^0$ production.

\begin{figure}[tb]
\begin{center}
\subfloat[\label{dd:a}]{
\includegraphics[scale=1,width=7.5cm]{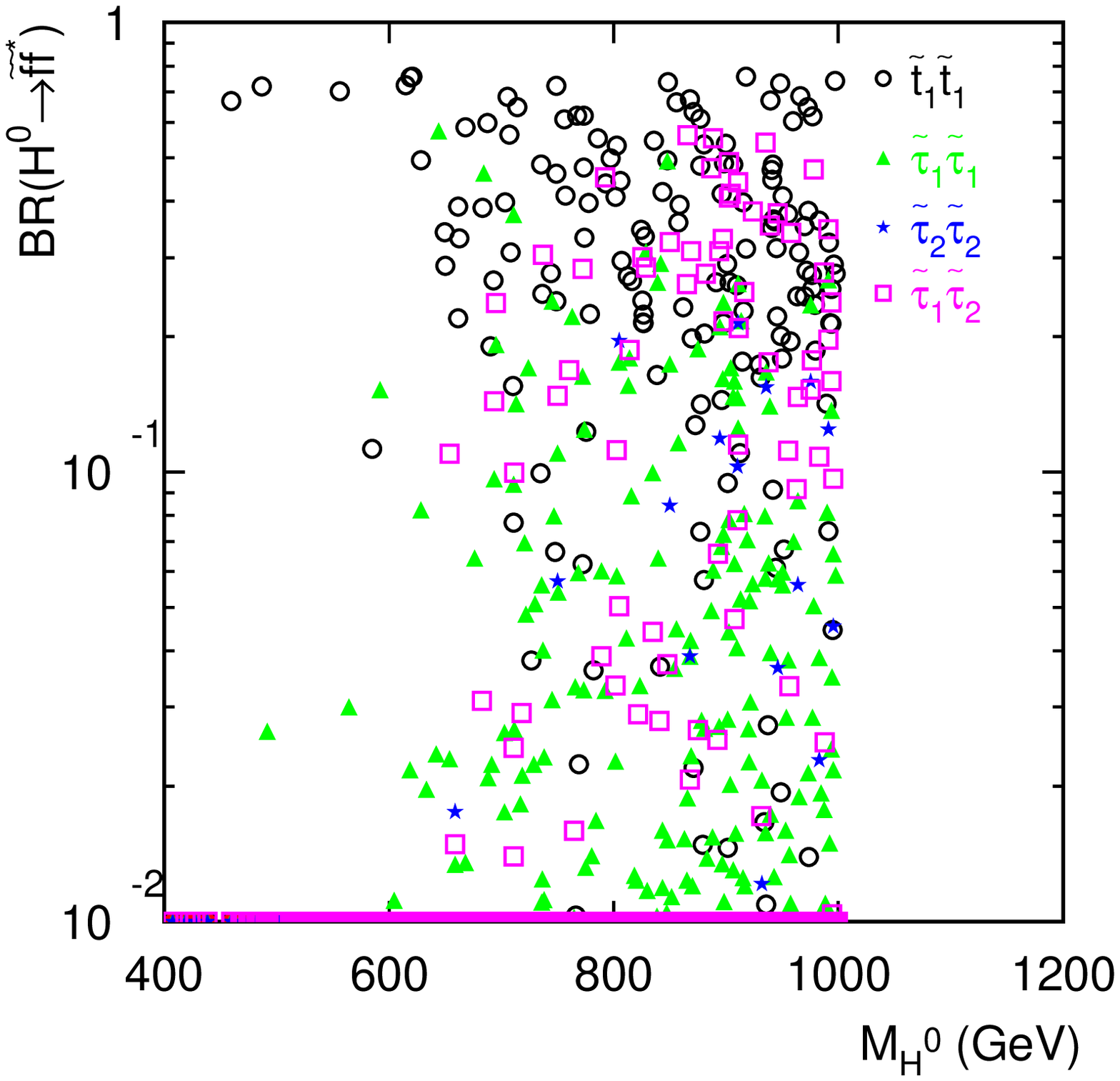}}
\subfloat[\label{dd:b}]{
\includegraphics[scale=1,width=7.5cm]{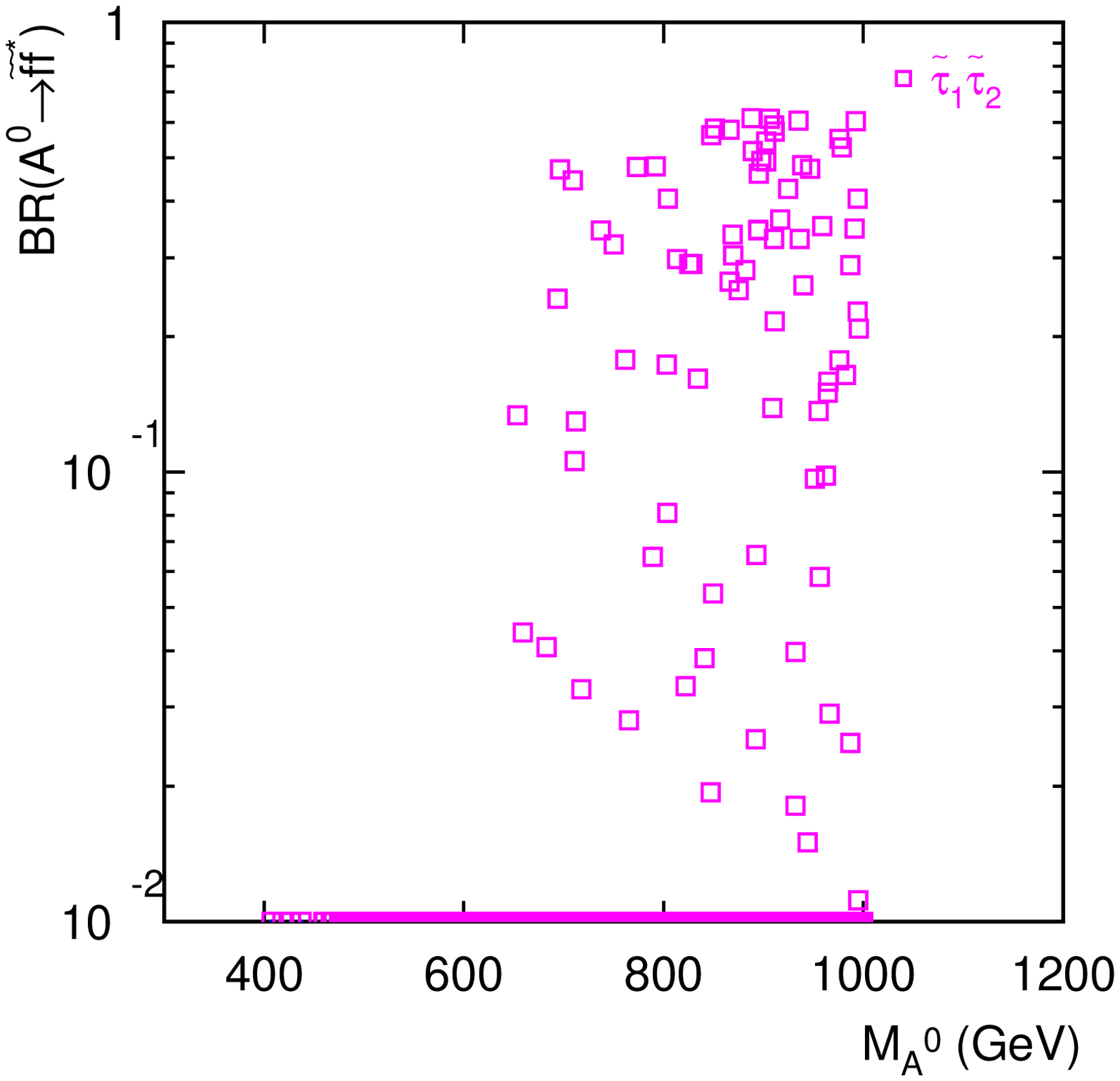}}\\
\subfloat[\label{dd:c}]{
\includegraphics[scale=1,width=7.5cm]{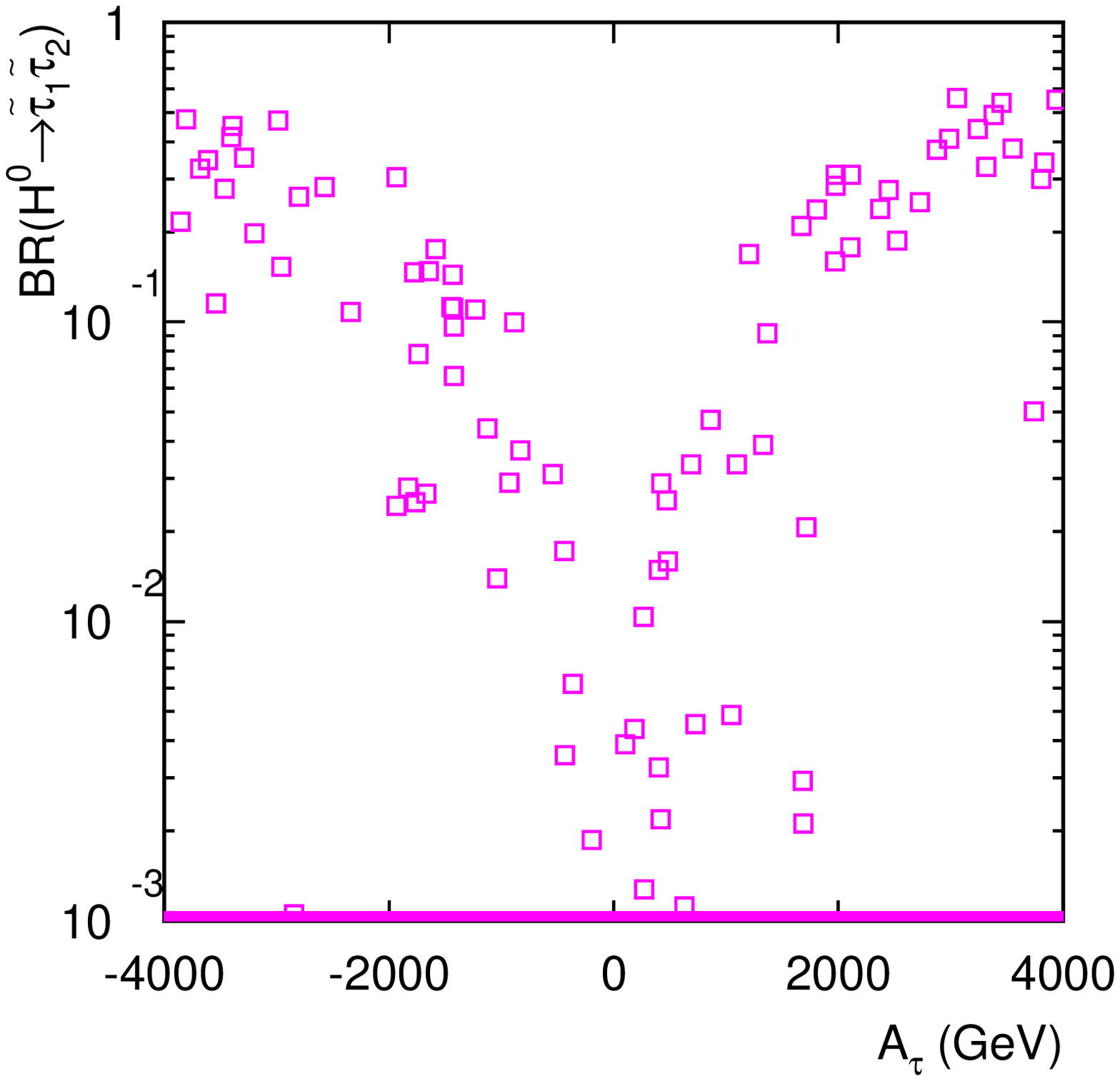}}
\subfloat[\label{dd:d}]{
\includegraphics[scale=1,width=7.5cm]{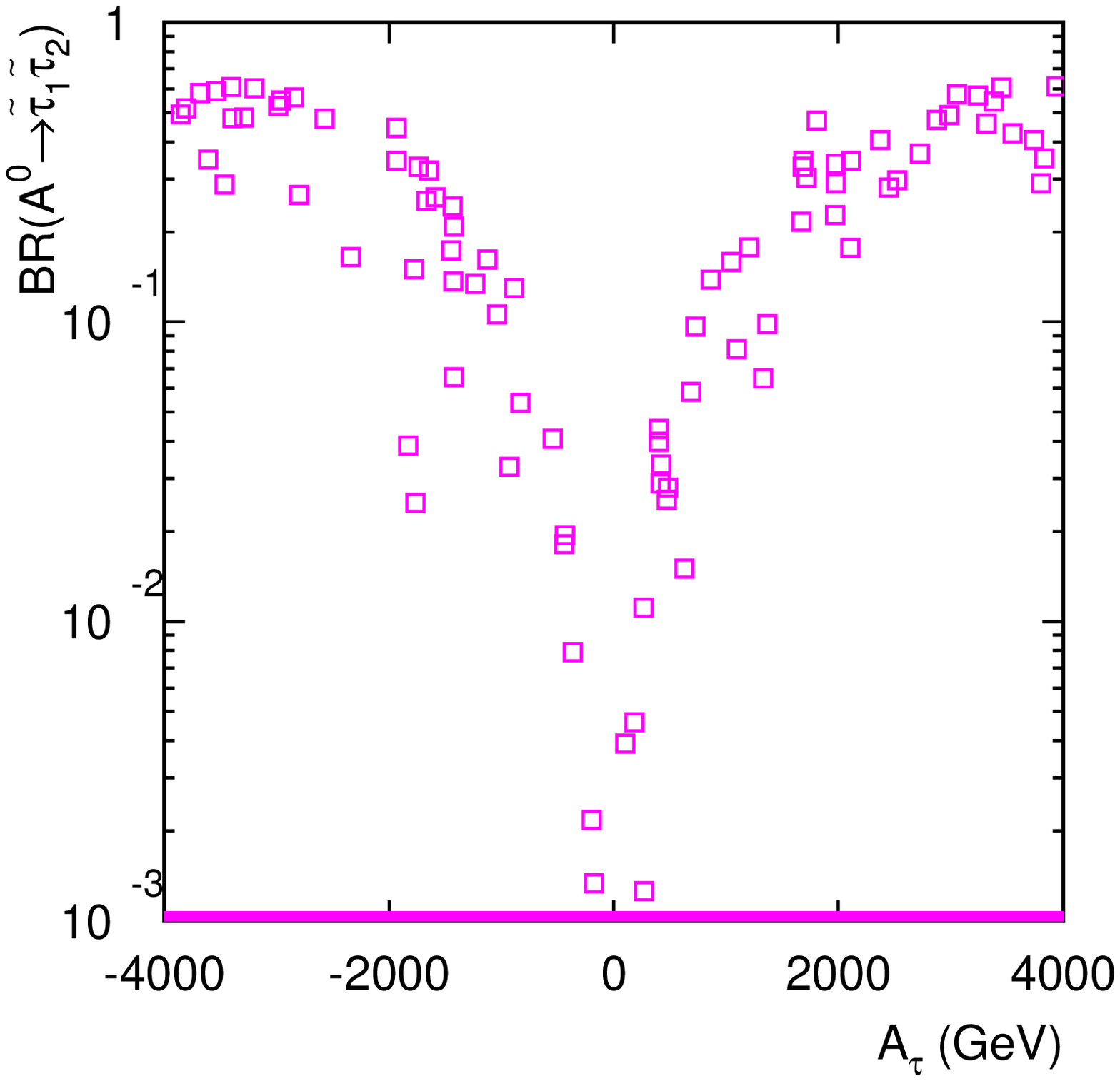}}
\end{center}
\caption{(a) $BR(H^0\to \tilde{f}\tilde{f}^\ast)$ vs. $m_{H^0}$,
(b) $BR(A^0\to \tilde{f}\tilde{f}^\ast)$ vs. $m_{A^0}$, (c) $BR(H^0\to \tilde{\tau}_1\tilde{\tau}_2)$ vs. $A_\tau$ and (d) $BR(A^0\to \tilde{\tau}_1\tilde{\tau}_2)$ vs. $A_\tau$.}
\label{sfsfa}
\end{figure}

\begin{figure}[tb]
\begin{center}
\subfloat[\label{dd:a}]{
\includegraphics[scale=1,width=8.5cm]{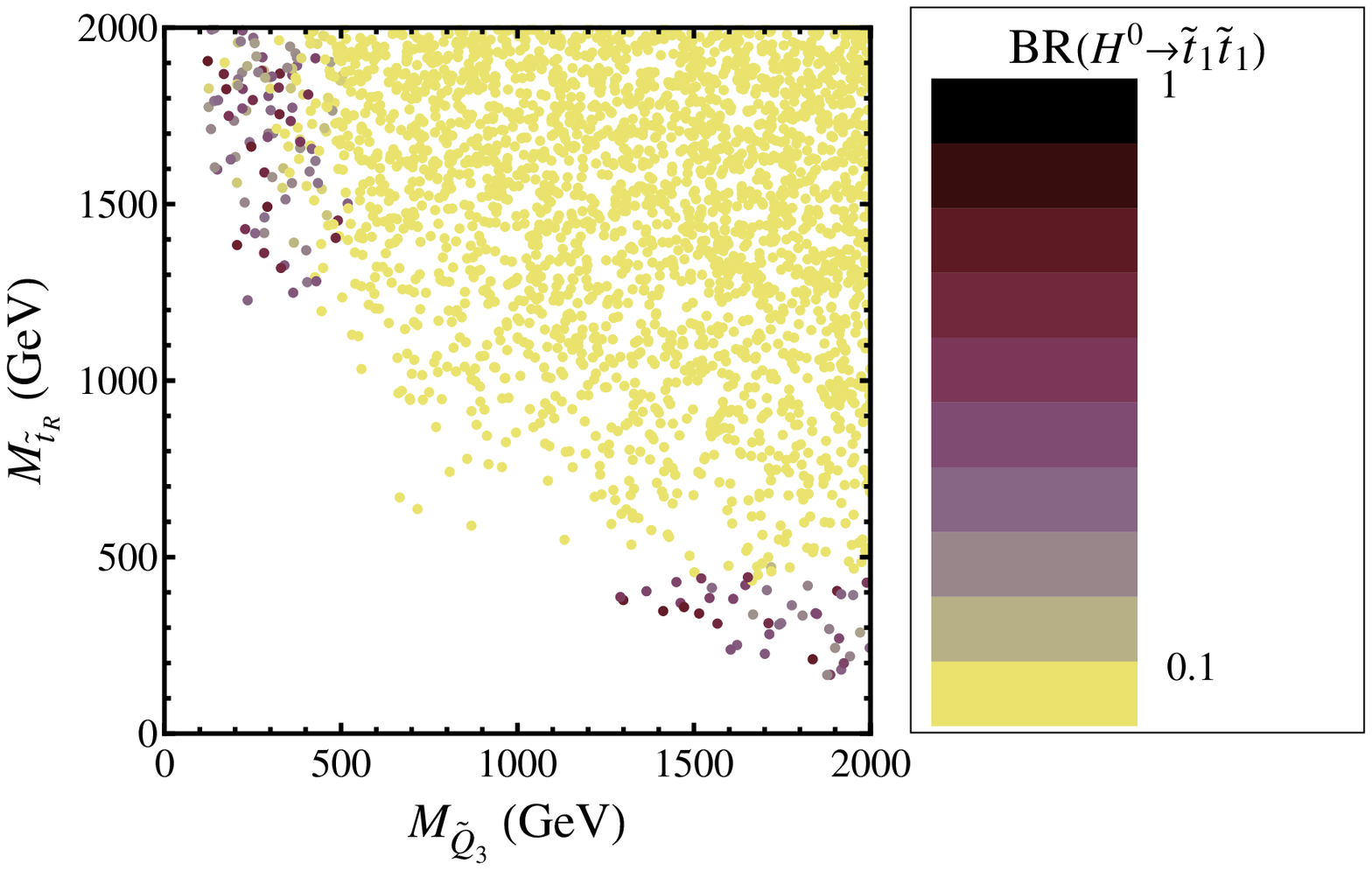}}
\subfloat[\label{dd:b}]{
\includegraphics[scale=1,width=8.5cm]{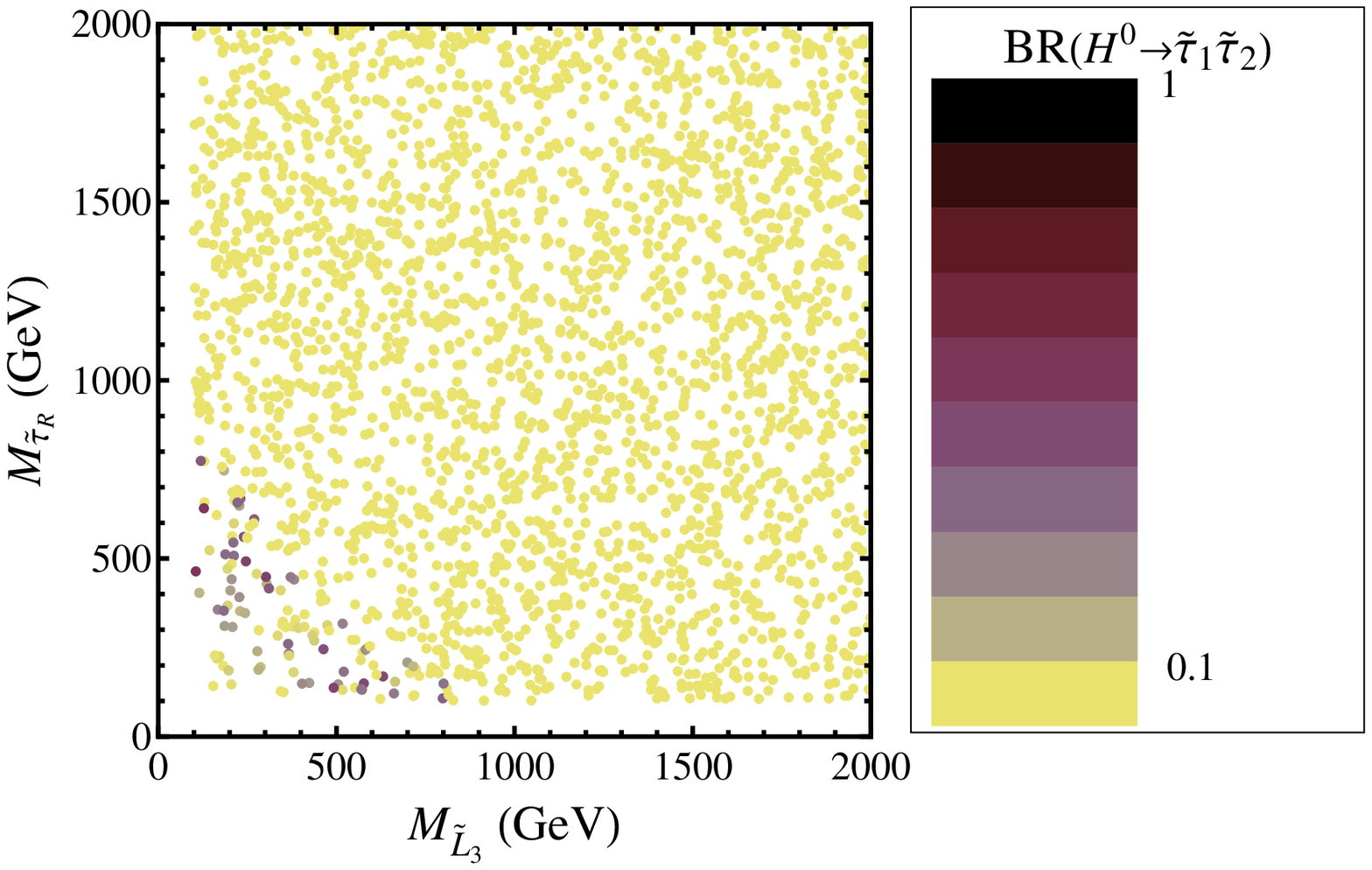}}
\end{center}
\caption{(a) $BR(H^0\to \tilde{t}_1\tilde{t}_1^\ast)$ in the plane of $M_{\tilde{t}_R}$ vs. $M_{\tilde{Q}_3}$ and (b) $BR(H^0\to \tilde{\tau}_1\tilde{\tau}_2^\ast+\tilde{\tau}_2\tilde{\tau}_1^\ast)$ in the plane of $M_{\tilde{\tau}_R}$ vs. $M_{\tilde{L}_3}$. The color scale gives the branching fraction of $H^0\to \tilde{f}\tilde{f}^\ast$ decay.} \label{sfsfb}
\end{figure}

Indicated by the fit to Higgs mass and signals, light sfermions also play important role in the heavy Higgs decay.
In the decoupling limit, the heavy neutral Higgses couplings to sfermion current eigenstates are given by~\cite{Djouadi:2005gj}
\begin{eqnarray}
C_{H^0\tilde{f}\tilde{f}}&=&\left(
                            \begin{array}{cc}
                              (I_f^{3L}-Q_fs_W^2)m_Z^2\sin2\beta+m_f^2r_1^f & {1\over 2}m_f(A_fr_1^f+\mu r_2^f) \\
                              {1\over 2}m_f(A_fr_1^f+\mu r_2^f) & Q_fs_W^2m_Z^2\sin2\beta+m_f^2r_1^f \\
                            \end{array}
                          \right),\\
C_{A^0\tilde{f}\tilde{f}}&=&\left(
                            \begin{array}{cc}
                              0 & -{1\over 2}m_f\left(A_f(\tan\beta)^{-2I_3^f}+\mu\right) \\
                              {1\over 2}m_f\left(A_f(\tan\beta)^{-2I_3^f}+\mu\right) & 0 \\
                            \end{array}
                          \right),
\end{eqnarray}
where $r_1^u=-\cot\beta$, $r_1^d=r_1^l=-\tan\beta$, $r_2^u=-1$ and $r_2^d=r_2^l=1$.
For CP-even Higgs $H^0$, these couplings contain term proportional to $m_f^2$ and thus get enhanced for the third generation sfermions.
%are thus potentially large since they involve terms $m_t^2\cot\beta$ when $\tan\beta$ is small in the
%stop case and, in the case of stau, there are terms $m_\tau^2\tan\beta$ that can be strongly
%enhanced for large values of $\tan\beta$.
The CP-odd Higgs $A^0$ only couples to $\tilde{f}_1\tilde{f}_2$ mixtures
with couplings $\propto m_f$. The stop decay mode for $A^0$ is then forbidden as at least one stop has to be very heavy to accommodate SM-like Higgs mass. Figs.~\ref{sfsfa} (a) and (b) show that the branching fraction of heavy Higgses decay into sfermions can be as large as 80\% for $H^0\to \tilde{t}_1\tilde{t}_1^\ast$ and 60\% for $H^0/A^0\to \tilde{\tau}_1\tilde{\tau}_2^\ast+\tilde{\tau}_1^\ast\tilde{\tau}_2$.
Moreover, with increasing $|A_\tau|$, both $H^0$ and $A^0$ have increasing branching ratio of $\tilde{\tau}_1\tilde{\tau}_2^\ast+\tilde{\tau}_1^\ast\tilde{\tau}_2$ decay mode~\cite{carlosstopstau}, as seen from Figs.~\ref{sfsfa} (c) and (d).
In Fig.~\ref{sfsfb} we display the dependence of heavy Higgs decay into light sfermions on SUSY soft masses. The decay $H^0\to \tilde{t}_1\tilde{t}_1^\ast$ is dominant for either $m_{\tilde{Q}_3}<500$ GeV, $m_{\tilde{t}_R}>1.2$ TeV or $m_{\tilde{Q}_3}>$ 1.2 TeV, $m_{\tilde{t}_R}<500$ GeV with only one light stop.
While $H^0\to \tilde{\tau}_1\tilde{\tau}_2^\ast+\tilde{\tau}_1^\ast\tilde{\tau}_2$ could be dominant for $m_{\tilde{L}_3},m_{\tilde{\tau}_R}<800$ GeV with two light staus.

\begin{figure}[tb]
\begin{center}
\subfloat[\label{dd:a}]{
\includegraphics[scale=1,width=7.5cm]{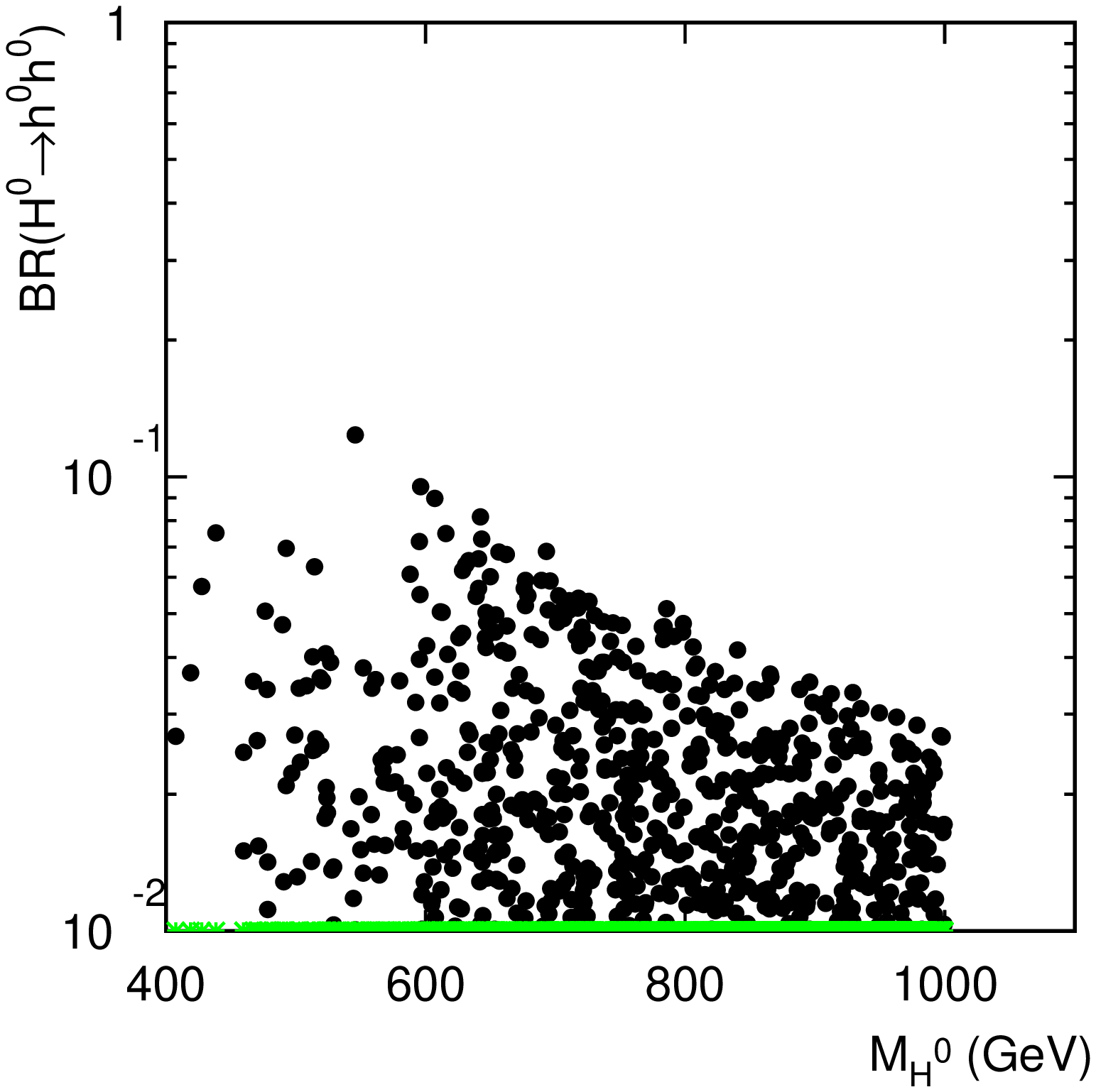}}
\subfloat[\label{dd:b}]{
\includegraphics[scale=1,width=7.5cm]{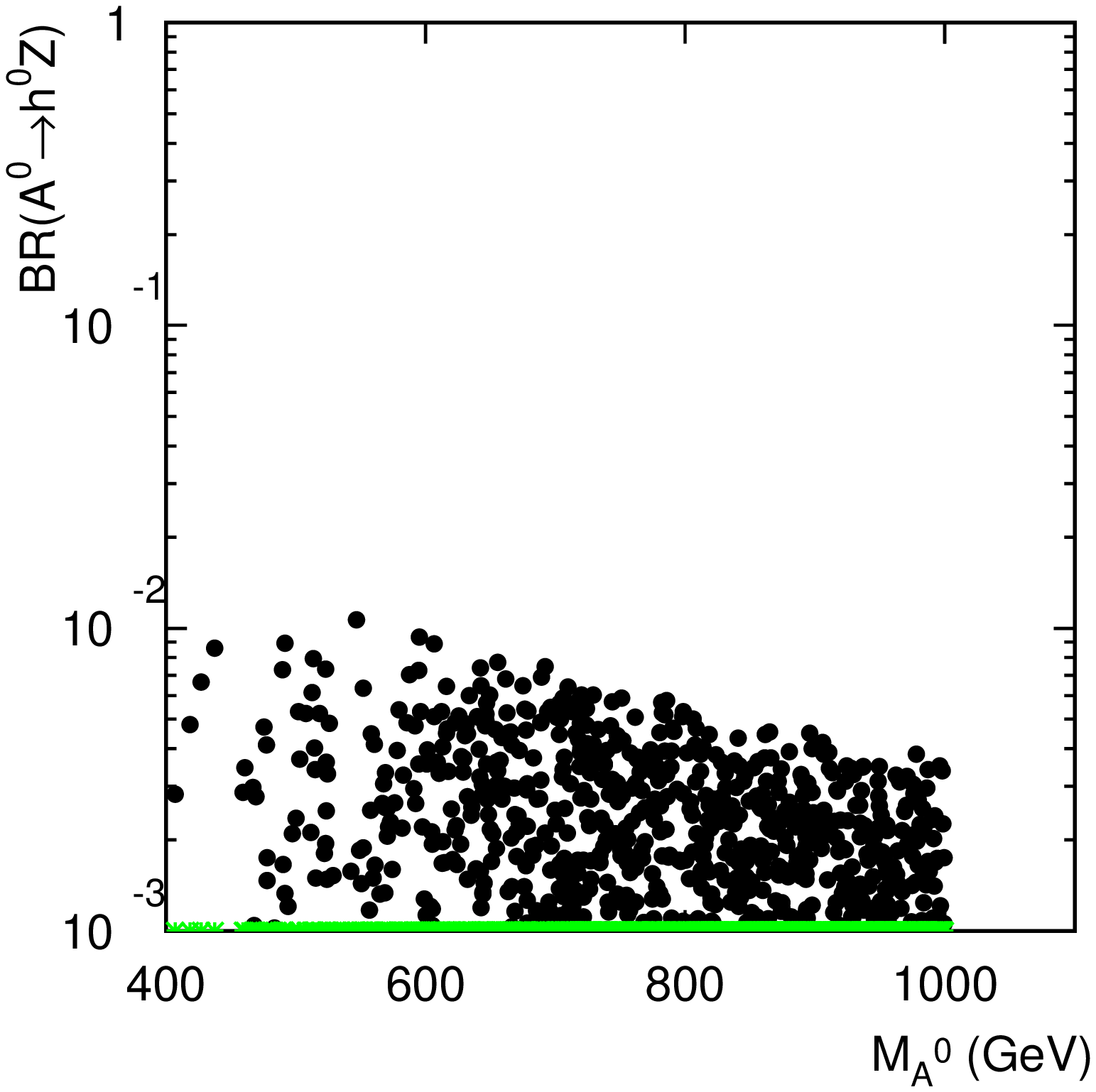}}
\end{center}
\caption{(a) $BR(H^0\to h^0h^0)$ vs. $m_{H^0}$ and (b) $BR(A^0\to h^0Z)$ vs. $m_{A^0}$.} \label{hv}
\end{figure}

The decays $H^0\to h^0h^0$ and $A^0\to h^0 Z$ are known to complement heavy Higgs searches at low values of $\tan\beta$ and intermediate $M_A$ masses~\cite{Arbey,hv}.
In the decoupling limit with $M_A>400$ GeV constrained by current Higgs searches, the corresponding partial decay widths are suppressed by $1/m_{H^0}$ and coupling $\cos^2(\beta-\alpha)\ll 1$, respectively. Their branching fractions are thus decreasing quickly with at most 10\% for $H^0\to h^0h^0$ and 1\% for $A^0\to h^0Z$ as seen in Fig.~\ref{hv}.

\begin{figure}[tb]
\begin{center}
\subfloat[\label{dd:a}]{
\includegraphics[scale=1,width=7.5cm]{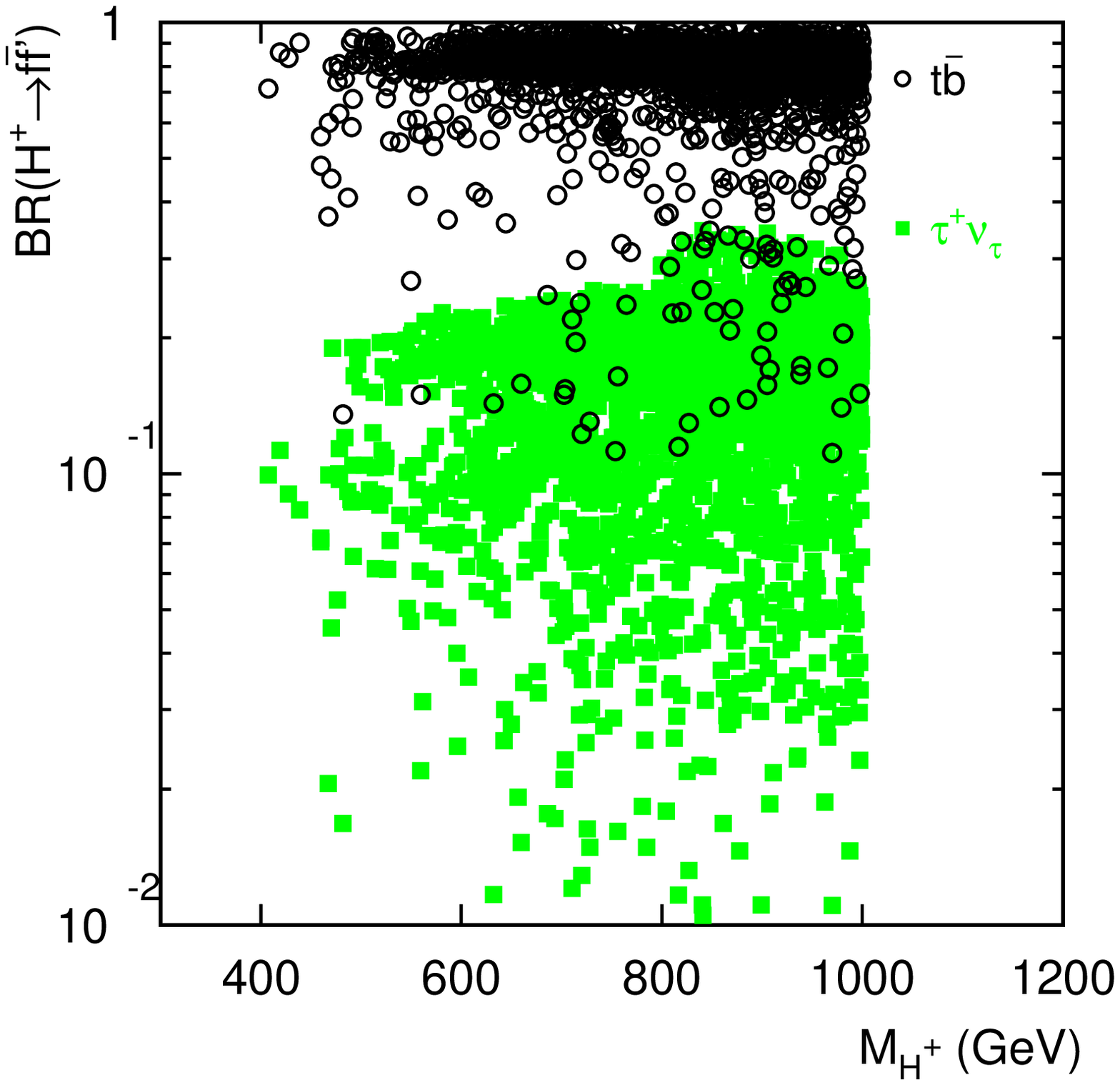}}
\subfloat[\label{dd:b}]{
\includegraphics[scale=1,width=7.5cm]{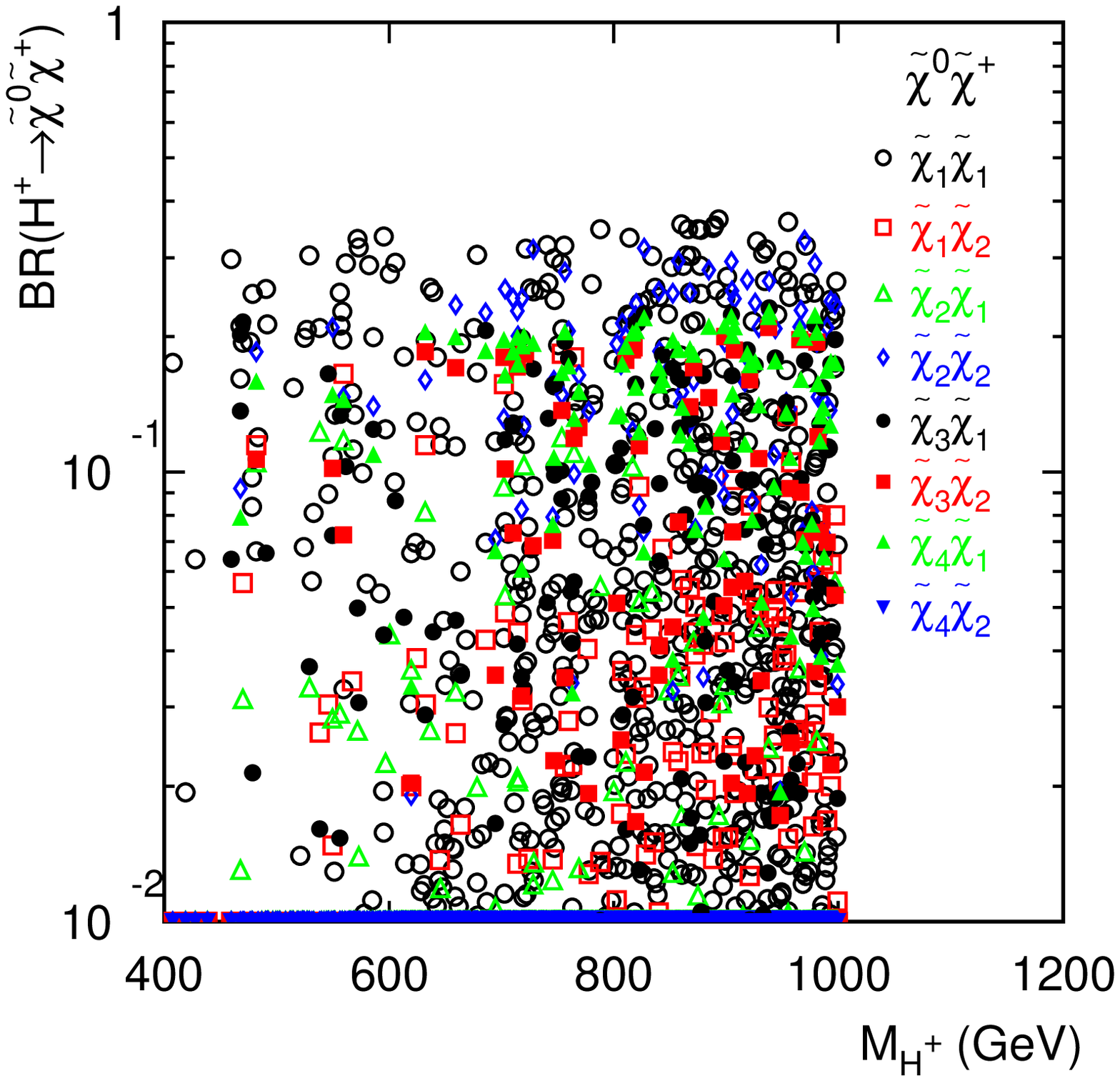}}\\
\subfloat[\label{dd:c}]{
\includegraphics[scale=1,width=7.5cm]{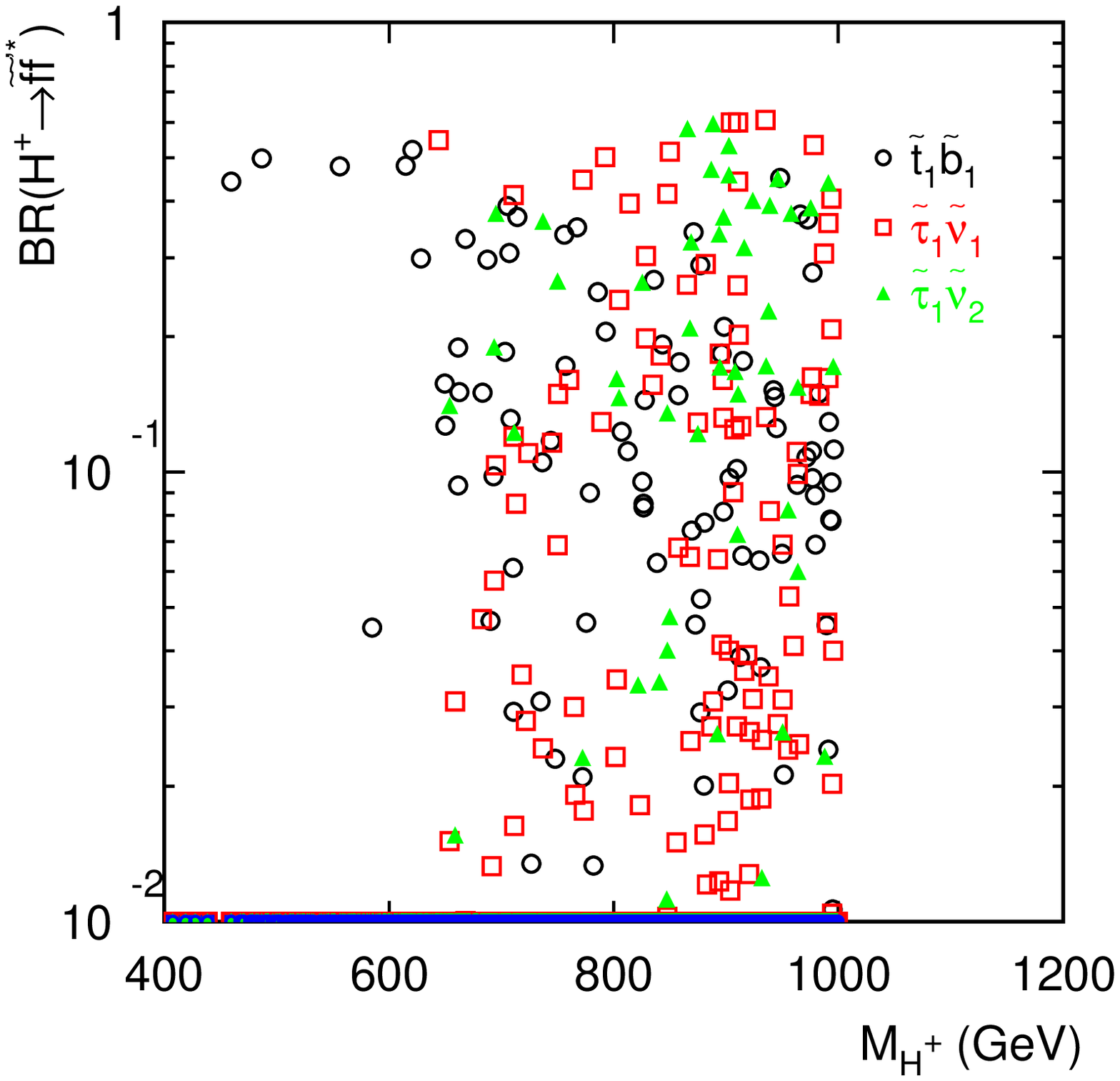}}
\end{center}
\caption{(a) $BR(H^+\to f\bar{f}')$ vs. $m_{H^+}$, (b) $BR(H^+\to \tilde{\chi}^0_i\tilde{\chi}^+_j)$ vs.
$m_{H^+}$ and (c) $BR(H^+\to \tilde{f}\tilde{f}'^\ast)$ vs. $m_{H^+}$.} \label{hpmf}
\end{figure}

Finally, we show the branching fraction of charged Higgs decay in Fig.~\ref{hpmf}. A majority of points give $BR(H^+\to t\bar{b})\sim 80\%$ and $BR(H^+\to \tau^+\nu_\tau)\sim 20\%$. The branching ratio of charged Higgs decay to light neutralino plus chargino and sfermions can be as large as 40\% and 60\%, respectively.

%\begin{figure}[tb]
%\begin{center}
%\includegraphics[scale=1,width=8cm]{../plots/densityHtanstst.eps}
%\includegraphics[scale=1,width=8cm]{../plots/densityHtanstasta12.eps}
%\end{center}
%\caption{$BR(H^0\to \tilde{f}\tilde{f}^\ast)$ in the plane of $\tan\beta$ vs. $M_{\tilde{Q}_3}$ or $\tan\beta$ vs. $M_{\tilde{L}_3}$. The color scale gives the %branching fraction of $H^0\to \tilde{f}\tilde{f}^\ast$ decay.} \label{sfsfc}
%\end{figure}

%\begin{figure}[tb]
%\begin{center}
%\includegraphics[scale=1,width=8cm]{../plots/densityHAtstst.eps}\\
%\includegraphics[scale=1,width=8cm]{../plots/densityHAtastasta.eps}
%\includegraphics[scale=1,width=8cm]{../plots/densityHmustasta.eps}
%\end{center}
%\caption{$BR(H^0\to \tilde{f}\tilde{f}^\ast)$ in the plane of $X_t/\sqrt{M_{\tilde{Q}_3}M_{\tilde{t}_R}}$ vs. $X_\tau$ or %$X_\tau/\sqrt{M_{\tilde{L}_3}M_{\tilde{\tau}_R}}$ vs. $A_\tau/\mu$. The color scale gives the branching fraction of $H^0\to \tilde{f}\tilde{f}^\ast$ decay.} %\label{sfsfd}
%\end{figure}

%%%%%%%%%%%%%%%%%%%%%%%%%%%%%%%%%%%%%%%%%%%%%%%%%%%%%%%%%55

%\begin{figure}[tb]
%\begin{center}
%\includegraphics[scale=1,width=8cm]{../plots/Hpmch.eps}
%\end{center}
%\caption{Left: $BR(H^+\to \tilde{\chi}^0_i\tilde{\chi}^+_j)$ vs.
%$m_{H^+}$. Black filled circle: $\tilde{\chi}^0_1\tilde{\chi}^+_1$,
%Red filled circle: $\tilde{\chi}^0_1\tilde{\chi}^+_2$, Green filled
%circle: $\tilde{\chi}^0_2\tilde{\chi}^+_1$, Blue filled circle:
%$\tilde{\chi}^0_2\tilde{\chi}^+_2$; Black open circle:
%$\tilde{\chi}^0_3\tilde{\chi}^+_1$, Red open circle:
%$\tilde{\chi}^0_3\tilde{\chi}^+_2$, Green open circle:
%$\tilde{\chi}^0_4\tilde{\chi}^+_1$, Blue open circle:
%$\tilde{\chi}^0_4\tilde{\chi}^+_2$.} \label{hpmch}
%\end{figure}

%\begin{figure}[tb]
%\begin{center}
%\includegraphics[scale=1,width=8cm]{../plots/Hpmsfsf.eps}
%\end{center}
%\caption{$BR(H^+\to \tilde{f}\tilde{f}'^\ast)$ vs. $m_{H^+}$. Black filled circle: $\tilde{t}_1\tilde{b}_1^\ast$,
%Black open circle:
%$\tilde{\tau}_1\tilde{\nu}_1^\ast$, Red open circle:
%$\tilde{\tau}_1\tilde{\nu}_2^\ast$.}
%\label{hpmsf}
%\end{figure}

%%%%%%%%%%%%%%%%%%%%%%%%%%%%%%%%%%%
\subsection{Future Heavy Higgs Search Sensitivity}

\begin{figure}[tb]
\begin{center}
\includegraphics[scale=1,width=8cm]{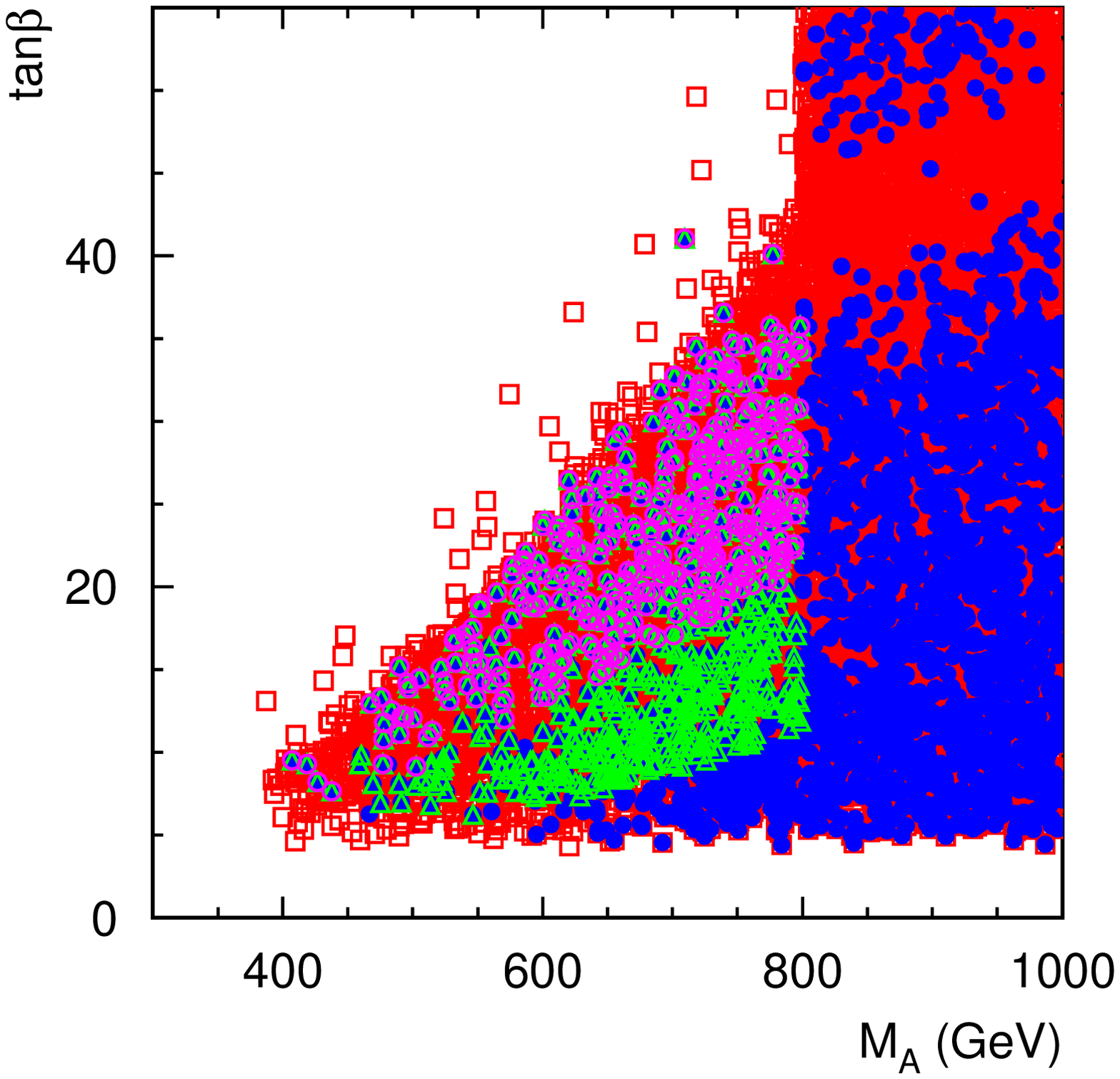}
\end{center}
\caption{Exclusion for $H^0/A^0\to \tau\tau$ mode in the plane of $\tan\beta$ vs. $m_A$ with $L=300$ fb$^{-1}$ (purple open circle) and 3000 fb$^{-1}$ (green open triangle), based on surviving region in Fig.~\ref{tbma}.} \label{excl}
\end{figure}

As one can see from previous subsection, SUSY effects could vary the $\tau\tau$ mode of heavy Higgs decay significantly. One has to consider the
variation of $\tau\tau$ exclusion limit given various SUSY decay products, for the small values of $\tan\beta$ in particular. We now improve measurement potential for the search of heavy MSSM Higgs decay into $\tau^+\tau^-$. Assuming the signal and background events go up by the same factor when the energy enhanced, we simply scale the signal sensitivity with $\sqrt{\sigma_{signal}\times L}$ based on the expected upper limit on the $\tau\tau$ channel~\cite{CMStau}, where $\sigma_{signal}=\sigma(gg,b\bar{b}\to H^0,A^0\to \tau^+\tau^-)$ at 14 TeV LHC and $L$ is the integrated luminosity. The extrapolation of excluded region for $\tau\tau$ mode at 14 TeV LHC is shown in Fig.~\ref{excl} with $L=300$ fb$^{-1}$ and 3000 fb$^{-1}$.
One can see that, in the plane of $\tan\beta-M_A$ with $M_A<800$ GeV, the $\tau\tau$ mode can only essentially exclude regime with $\tan\beta>20$ for $L=300$ fb$^{-1}$ and $\tan\beta>15$ for $L=3000$ fb$^{-1}$.

%%%%%%%%%%%%%%%%%%%%%%%%%%%%%%%%%%%%%%%%%%%%%%%%%%%%%%%%%%%%%%%%%%%%%%%%
\section{Conclusions}
%%%%%%%%%%%%%%%%%%%%%%%%%%%%%%%%%%%%%%%%%%%%%%%%%%%%%%%%%%%%%%%%%%%%%%%%

The decoupling limit in the MSSM Higgs sector is the most likely scenario in light of the Higgs discovery.
This scenario is further constrained by MSSM Higgs search bounds and flavor measurements. We performed a comprehensive scan of MSSM parameter space and updated the constraints on the decoupling MSSM Higgs sector in terms of 8 TeV data. The light SUSY spectrum in charge of SM-like Higgs mass and signal excesses was discussed. We highlighted the effect of light SUSY spectrum in the heavy neutral Higgs decay in the decoupling limit. We found that
the measured Higgs mass window and current Higgs search data push $m_A$ to at least 400 GeV. Further $b$ rare decays do not put more stringent constraints on the surviving region.
%allow the whole region of $m_A>400$ GeV and $5<\tan\beta<40$.
The chargino and neutralino decay mode can reach at most 40\% and 20\% branching ratio, respectively. In particular, the invisible decay $BR(H^0(A^0)\to \tilde{\chi}^0_1\tilde{\chi}^0_1)$ increases with increasing Bino LSP mass and sits between 10\%-15\% (20\%) for $30<m_{\tilde{\chi}^0_1}<100$ GeV. The branching fraction of dominant heavy Higgses decay into sfermions can be as large as 80\% for $H^0\to \tilde{t}_1\tilde{t}_1^\ast$ and 60\% for $H^0/A^0\to \tilde{\tau}_1\tilde{\tau}_2^\ast+\tilde{\tau}_1^\ast\tilde{\tau}_2$. $H^0\to h^0h^0$ and $A^0\to h^0Z$ have the branching fraction less than 10\% and 1\%, respectively, for $m_A>400$ GeV. The branching ratio of charged Higgs decay to neutralino plus chargino and sfermions can be as large as 40\% and 60\%, respectively. Moreover, these dominant SUSY products alter the normal heavy Higgs decay modes dramatically, in particular for small $\tan\beta$ region. We extrapolated the exclusion limit of leading MSSM Higgs search channel, namely $gg,b\bar{b}\to H^0, A^0\to \tau^+\tau^-$, to center-of-mass energy of 14 TeV with high luminosities at the LHC based on surviving region and exceptions of dominant SUSY decay channels. It turns out that the $\tau\tau$ mode can essentially exclude regime with $\tan\beta>20$ for $L=300$ fb$^{-1}$ and $\tan\beta>15$ for $L=3000$ fb$^{-1}$.

%%%%%%%%%%%%%%%%%%%%%%%%%%
\subsection*{Acknowledgment}
%%%%%%%%%%%%%%%%%%%%%%%%%%
We would like to thank Marcela Carena and Ian Lewis for useful discussions.
This work was supported in part by the Australian Research Council.
%This work is supported by the DOE under grant No. DE-FG02-91ER40626.
%This work used the Extreme Science and Engineering Discovery Environment (XSEDE), which is
%supported by National Science Foundation grant number OCI-1053575.

%%%%%%%%%%%%%%%%%%%%%%%%%%%%%%%%%%%%%%%%%%%%%%%%%%%%%%%%%%%%%%%%%%%%%%%%%
%\appendix

%%%%%%%%%%%%%%%%%%%%%%%%%%%%%%%%%%%%%%%%%%%%%%%
%%%%%%%%%%%%%%%%%%%%%%%%%%%%%%%%%%%%%%%%%%%%%%%%%%%%%%%%%%%%%%%%%%%%
%%%%%%%%%%%%%%%%%%%%%%%%%%%%%


\begin{thebibliography}{000}
%%%%%%%%%%%%%%%%%%%%%%%%%%%%%%%%%%

\bibitem{:2012gk}
  G.~Aad {\it et al.}  [ATLAS Collaboration],
  %``Observation of a new particle in the search for the Standard Model Higgs boson with the ATLAS detector at the LHC,''
  Phys.\ Lett.\ B {\bf 716}, 1 (2012)
  [arXiv:1207.7214 [hep-ex]];
  %%CITATION = ARXIV:1207.7214;%%
%%\cite{:2012gu}
%\bibitem{:2012gu}
  S.~Chatrchyan {\it et al.}  [CMS Collaboration],
  %``Observation of a new boson at a mass of 125 GeV with the CMS experiment at the LHC,''
  Phys.\ Lett.\ B {\bf 716}, 30 (2012)
  [arXiv:1207.7235 [hep-ex]].
  %%CITATION = ARXIV:1207.7235;%%


%\cite{Gunion:1989we}
\bibitem{Gunion:1989we}
  J.~F.~Gunion, H.~E.~Haber, G.~L.~Kane and S.~Dawson,
  %``The Higgs Hunter's Guide,''
  Front.\ Phys.\  {\bf 80}, 1 (2000);
  %%CITATION = FRPHA,80,1;%%
%\cite{Gunion:1984yn}
%
%\bibitem{Gunion:1984yn}
  J.~F.~Gunion and H.~E.~Haber,
  %``Higgs Bosons in Supersymmetric Models. 1.,''
  Nucl.\ Phys.\ B {\bf 272}, 1 (1986)
  [Erratum-ibid.\ B {\bf 402}, 567 (1993)].
  %%CITATION = NUPHA,B272,1;%%

%\cite{Djouadi:2005gj}
\bibitem{Djouadi:2005gj}
  A.~Djouadi,
  %``The Anatomy of electro-weak symmetry breaking. II. The Higgs bosons in the minimal supersymmetric model,''
  Phys.\ Rept.\  {\bf 459}, 1 (2008)
  [hep-ph/0503173].
  %%CITATION = HEP-PH/0503173;%%

\bibitem{Christensen:2012ei}
  N.~D.~Christensen, T.~Han and S.~Su,
  %``MSSM Higgs Bosons at The LHC,''
  Phys.\ Rev.\ D {\bf 85}, 115018 (2012)
  [arXiv:1203.3207 [hep-ph]].
  %%CITATION = ARXIV:1203.3207;%%



\bibitem{Haber:1994mt}
H.~E. Haber,
%{\it {Nonminimal Higgs sectors: The Decoupling limit and its
 % phenomenological implications}},
%  \href{http://xxx.lanl.gov/abs/hep-ph/9501320}{{\tt hep-ph/9501320}}.
hep-ph/9501320.

%\cite{Haber:1995be}
\bibitem{Haber:1995be}
  H.~E.~Haber,
  %``Challenges for nonminimal Higgs searches at future colliders,''
  hep-ph/9505240.

\bibitem{Christensen:2012si}
  N.~D.~Christensen, T.~Han and T.~Li,
  %``Pair Production of MSSM Higgs Bosons in the Non-decoupling Region at the LHC,''
  Phys.\ Rev.\ D {\bf 86}, 074003 (2012)
  [arXiv:1206.5816 [hep-ph]].
  %%CITATION = ARXIV:1206.5816;%%
  %8 citations counted in INSPIRE as of 22 May 2013

\bibitem{nondecoupling}
 T.~Han, T.~Li, S.~Su and L.~-T.~Wang,
  %``Non-Decoupling MSSM Higgs Sector and Light Superpartners,''
  arXiv:1306.3229 [hep-ph].
  %%CITATION = ARXIV:1306.3229;%%
  %1 citations counted in INSPIRE as of 17 Sep 2013

\bibitem{Arbey}
A.~Arbey, M.~Battaglia and F.~Mahmoudi,
  %``Supersymmetric Heavy Higgs Bosons at the LHC,''
  Phys.\ Rev.\ D {\bf 88}, 015007 (2013)
  [arXiv:1303.7450 [hep-ph]].
  %%CITATION = ARXIV:1303.7450;%%
  %12 citations counted in INSPIRE as of 17 Sep 2013


\bibitem{CMStau}
S.~Chatrchyan {\it et al.}  [CMS Collaboration], CMS PAS HIG-12-050.


\bibitem{Ian}
I.~M.~Lewis,
  %``Closing the Wedge with 300 fb^-1 and 3000 fb^-1 at the LHC: A Snowmass White Paper,''
  arXiv:1308.1742 [hep-ph].
  %%CITATION = ARXIV:1308.1742;%%
  %1 citations counted in INSPIRE as of 17 Sep 2013

\bibitem{carlos}
M.~Carena, S.~Gori, N.~R.~Shah, C.~E.~M.~Wagner and L.~-T.~Wang,
  %``Light Stau Phenomenology and the Higgs $\gamma\gamma$ Rate,''
  JHEP {\bf 1207}, 175 (2012)
  [arXiv:1205.5842 [hep-ph]].
  %%CITATION = ARXIV:1205.5842;%%
  %125 citations counted in INSPIRE as of 03 Sep 2013

\bibitem{taodm}
T.~Han, Z.~Liu and A.~Natarajan,
  %``Dark Matter and Higgs Bosons in the MSSM,''
  arXiv:1303.3040 [hep-ph].
  %%CITATION = ARXIV:1303.3040;%%
  %15 citations counted in INSPIRE as of 03 Sep 2013

  %%%%%%%%%%%%%%%%%%%%%%%%%%%%%
  %                               FeynHiggs
  %%%%%%%%%%%%%%%%%%%%%%%%%%%%%
    %\cite{Degrassi:2002fi}
\bibitem{Degrassi:2002fi}
  G.~Degrassi, S.~Heinemeyer, W.~Hollik, P.~Slavich and G.~Weiglein,
  %``Towards high precision predictions for the MSSM Higgs sector,''
  Eur.\ Phys.\ J.\ C {\bf 28}, 133 (2003)
  [hep-ph/0212020].
  %%CITATION = HEP-PH/0212020;%%

  %\cite{Heinemeyer:1998np}
\bibitem{Heinemeyer:1998np}
  S.~Heinemeyer, W.~Hollik and G.~Weiglein,
  %``The Masses of the neutral CP - even Higgs bosons in the MSSM: Accurate analysis at the two loop level,''
  Eur.\ Phys.\ J.\ C {\bf 9}, 343 (1999)
  [hep-ph/9812472].
  %%CITATION = HEP-PH/9812472;%%

% b-> s gamma, chargino loop effects
  %\cite{Frank:2006yh}
\bibitem{Frank:2006yh}
  M.~Frank, T.~Hahn, S.~Heinemeyer, W.~Hollik, H.~Rzehak and G.~Weiglein,
  %``The Higgs Boson Masses and Mixings of the Complex MSSM in the Feynman-Diagrammatic Approach,''
  JHEP {\bf 0702}, 047 (2007)  [hep-ph/0611326], and references therein.
  %%CITATION = HEP-PH/0611326;%%

  %\cite{Heinemeyer:1998yj}
\bibitem{Heinemeyer:1998yj}
  S.~Heinemeyer, W.~Hollik and G.~Weiglein,
  %``FeynHiggs: A Program for the calculation of the masses of the neutral CP even Higgs bosons in the MSSM,''
  Comput.\ Phys.\ Commun.\  {\bf 124}, 76 (2000)  [hep-ph/9812320], and references therein.
  %%CITATION = HEP-PH/9812320;%%

  %%%%%%%%%%%%%%%%%%%%%%%%%%%%%
  %                               HiggsBound
  %%%%%%%%%%%%%%%%%%%%%%%%%%%%%
  %\cite{Bechtle:2008jh}
\bibitem{Bechtle:2008jh}
  P.~Bechtle, O.~Brein, S.~Heinemeyer, G.~Weiglein and K.~E.~Williams,
  %``HiggsBounds: Confronting Arbitrary Higgs Sectors with Exclusion Bounds from LEP and the Tevatron,''
  Comput.\ Phys.\ Commun.\  {\bf 181}, 138 (2010)
  [arXiv:0811.4169 [hep-ph]], and references therein;
  %%CITATION = ARXIV:0811.4169;%%
%
  %\cite{Bechtle:2011sb}
%\bibitem{Bechtle:2011sb}
%  P.~Bechtle, O.~Brein, S.~Heinemeyer, G.~Weiglein and K.~E.~Williams,
  %``HiggsBounds 2.0.0: Confronting Neutral and Charged Higgs Sector Predictions with Exclusion Bounds from LEP and the Tevatron,''
{\it ibid.},  Comput.\ Phys.\ Commun.\  {\bf 182}, 2605 (2011)
  [arXiv:1102.1898 [hep-ph]], and references therein.
  %%CITATION = ARXIV:1102.1898;%%

  %%%%%%%%%%%%%%%%%%%%%%%%%%%%%
  %                               LEP
  %%%%%%%%%%%%%%%%%%%%%%%%%%%%%
  \bibitem{LEP2H}
Combined results from the LEP2 experiments, Phys.~Lett.~B{\bf 565}, 61 (2003).


  %%%%%%%%%%%%%%%%%%%%%%%%%%%%%
  %                               Tevatron
  %%%%%%%%%%%%%%%%%%%%%%%%%%%%%
\bibitem{CDFD0}
  %\cite{Aaltonen:2009ke}
%\bibitem{Aaltonen:2009ke}
  T.~Aaltonen {\it et al.}  [CDF Collaboration],
  %``Search for charged Higgs bosons in decays of top quarks in p anti-p collisions at s**(1/2) = 1.96 TeV,''
  Phys.\ Rev.\ Lett.\  {\bf 103}, 101803 (2009)
  [arXiv:0907.1269 [hep-ex]];
  %%CITATION = ARXIV:0907.1269;%%
  %\cite{:2009zh}
%
%\bibitem{2009zh}
  V.~M.~Abazov {\it et al.}  [D0 Collaboration],
  %``Search for charged Higgs bosons in top quark decays,''
  Phys.\ Lett.\ B {\bf 682}, 278 (2009)
  [arXiv:0908.1811 [hep-ex]].
  %%CITATION = ARXIV:0908.1811;%%

  \bibitem{Amhis:2012bh}
  Y.~Amhis {\it et al.}  [Heavy Flavor Averaging Group Collaboration],
  %``Averages of B-Hadron, C-Hadron, and tau-lepton properties as of early 2012,''
  arXiv:1207.1158 [hep-ex].
  %%CITATION = ARXIV:1207.1158;%%
  %160 citations counted in INSPIRE as of 22 May 2013


\bibitem{Aaij:2012nna}
  RAaij {\it et al.}  [LHCb Collaboration],
  %``First evidence for the decay Bs -> mu+ mu-,''
  Phys.\ Rev.\ Lett.\  {\bf 110}, 021801 (2013)
  [arXiv:1211.2674 [Unknown]].
  %%CITATION = ARXIV:1211.2674;%%
  %113 citations counted in INSPIRE as of 22 May 2013

\bibitem{Misiak:2006zs}
M.~Misiak, H.~M.~Asatrian, K.~Bieri, M.~Czakon, A.~Czarnecki, T.~Ewerth, A.~Ferroglia and P.~Gambino {\it et al.},
  %``Estimate of B(anti-B ---> X(s) gamma) at O(alpha(s)**2),''
  Phys.\ Rev.\ Lett.\  {\bf 98}, 022002 (2007)
  [hep-ph/0609232];
  %%CITATION = HEP-PH/0609232;%%
  %591 citations counted in INSPIRE as of 22 May 2013
T.~Becher and M.~Neubert,
  %``Analysis of Br(anti-B ---> X(s gamma)) at NNLO with a cut on photon energy,''
  Phys.\ Rev.\ Lett.\  {\bf 98}, 022003 (2007)
  [hep-ph/0610067].
  %%CITATION = HEP-PH/0610067;%%
  %127 citations counted in INSPIRE as of 22 May 2013


\bibitem{bsmumuSM}
K.~S.~Babu and C.~F.~Kolda,
  %``Higgs mediated $B^0 \to \mu^{+} \mu^{-}$ in minimal supersymmetry,''
  Phys.\ Rev.\ Lett.\  {\bf 84}, 228 (2000)
  [hep-ph/9909476];
  %%CITATION = HEP-PH/9909476;%%
  %393 citations counted in INSPIRE as of 21 May 2013
A.~J.~Buras, J.~Girrbach, D.~Guadagnoli and G.~Isidori,
  %``On the Standard Model prediction for BR(B{s,d} to mu+ mu-),''
  Eur.\ Phys.\ J.\ C {\bf 72}, 2172 (2012)
  [arXiv:1208.0934 [hep-ph]].
  %%CITATION = ARXIV:1208.0934;%%
  %62 citations counted in INSPIRE as of 22 May 2013

\bibitem{Misiak:2006ab}
  M.~Misiak and M.~Steinhauser,
  %``NNLO QCD corrections to the anti-B ---> X(s) gamma matrix elements using interpolation in m(c),''
  Nucl.\ Phys.\ B {\bf 764}, 62 (2007)
  [hep-ph/0609241].
  %%CITATION = HEP-PH/0609241;%%
  %199 citations counted in INSPIRE as of 22 May 2013

\bibitem{Lees:2012xj}
  J.~P.~Lees {\it et al.}  [BaBar Collaboration],
  %``Evidence for an excess of $\bar{B} \to D^{(*)} \tau^-\bar{\nu}_\tau$ decays,''
  Phys.\ Rev.\ Lett.\  {\bf 109}, 101802 (2012)
  [arXiv:1205.5442 [hep-ex]].
  %%CITATION = ARXIV:1205.5442;%%
  %69 citations counted in INSPIRE as of 22 May 2013

\bibitem{superiso}
F.~Mahmoudi,
  %``SuperIso: A Program for calculating the isospin asymmetry of B ---> K* gamma in the MSSM,''
  Comput.\ Phys.\ Commun.\  {\bf 178}, 745 (2008)
  [arXiv:0710.2067 [hep-ph]];
  %%CITATION = ARXIV:0710.2067;%%
  %109 citations counted in INSPIRE as of 22 May 2013
F.~Mahmoudi,
  %``SuperIso v2.3: A Program for calculating flavor physics observables in Supersymmetry,''
  Comput.\ Phys.\ Commun.\  {\bf 180}, 1579 (2009)
  [arXiv:0808.3144 [hep-ph]];
  %%CITATION = ARXIV:0808.3144;%%
  %127 citations counted in INSPIRE as of 22 May 2013
F.~Mahmoudi,
  %``SuperIso v3.0, flavor physics observables calculations: Extension to NMSSM,''
  Comput.\ Phys.\ Commun.\  {\bf 180}, 1718 (2009).
  %%CITATION = CPHCB,180,1718;%%
  %26 citations counted in INSPIRE as of 22 May 2013



\bibitem{Carena:1995wu}
  M.~S.~Carena, M.~Quiros and C.~E.~M.~Wagner,
  %``Effective potential methods and the Higgs mass spectrum in the MSSM,''
  Nucl.\ Phys.\  B {\bf 461}, 407 (1996)
  [arXiv:hep-ph/9508343];
  %%CITATION = NUPHA,B461,407;%%
%
%\cite{Carena:1995bx}
%\bibitem{Carena:1995bx}
  M.~S.~Carena, J.~R.~Espinosa, M.~Quiros and C.~E.~M.~Wagner,
  %``Analytical expressions for radiatively corrected Higgs masses and couplings
  %in the MSSM,''
  Phys.\ Lett.\  B {\bf 355}, 209 (1995)
  [arXiv:hep-ph/9504316].
  %%CITATION = PHLTA,B355,209;%%


\bibitem{carlosstopstau}
M.~Carena, S.~Gori, N.~R.~Shah, C.~E.~M.~Wagner and L.~-T.~Wang,
  %``Light Stops, Light Staus and the 125 GeV Higgs,''
  JHEP {\bf 1308}, 087 (2013)
  [arXiv:1303.4414 [hep-ph]].
  %%CITATION = ARXIV:1303.4414;%%
  %10 citations counted in INSPIRE as of 17 Sep 2013



  \bibitem{Heinemeyer}
P.~Bechtle, S.~Heinemeyer, O.~Stal, T.~Stefaniak, G.~Weiglein and L.~Zeune,
  %``MSSM Interpretations of the LHC Discovery: Light or Heavy Higgs?,''
  Eur.\  Phys.\  J.\  C {\bf 73:2354} (2013)
  [arXiv:1211.1955 [hep-ph]].
  %%CITATION = ARXIV:1211.1955;%%
  %22 citations counted in INSPIRE as of 21 May 2013


\bibitem{carlosstau}
M.~Carena, S.~Gori, N.~R.~Shah and C.~E.~M.~Wagner,
  %``A 125 GeV SM-like Higgs in the MSSM and the $\gamma \gamma$ rate,''
  JHEP {\bf 1203}, 014 (2012)
  [arXiv:1112.3336 [hep-ph]].
  %;
  %%CITATION = ARXIV:1112.3336;%%
  %226 citations counted in INSPIRE as of 04 Sep 2013
M.~Carena, S.~Gori, N.~R.~Shah, C.~E.~M.~Wagner and L.~-T.~Wang,
  %``Light Stau Phenomenology and the Higgs $\gamma\gamma$ Rate,''
  JHEP {\bf 1207}, 175 (2012)
  [arXiv:1205.5842 [hep-ph]].
  %%CITATION = ARXIV:1205.5842;%%
  %125 citations counted in INSPIRE as of 04 Sep 2013


\bibitem{stopamp}
K.~Blum, R.~T.~D'Agnolo and J.~Fan,
  %``Natural SUSY Predicts: Higgs Couplings,''
  JHEP {\bf 1301}, 057 (2013)
  [arXiv:1206.5303 [hep-ph]];
  %%CITATION = ARXIV:1206.5303;%%
  %43 citations counted in INSPIRE as of 17 Sep 2013
M.~R.~Buckley and D.~Hooper,
  %``Are There Hints of Light Stops in Recent Higgs Search Results?,''
  Phys.\ Rev.\ D {\bf 86}, 075008 (2012)
  [arXiv:1207.1445 [hep-ph]];
  %%CITATION = ARXIV:1207.1445;%%
  %55 citations counted in INSPIRE as of 17 Sep 2013
J.~R.~Espinosa, C.~Grojean, V.~Sanz and M.~Trott,
  %``NSUSY fits,''
  JHEP {\bf 1212}, 077 (2012)
  [arXiv:1207.7355 [hep-ph]].
  %%CITATION = ARXIV:1207.7355;%%
  %39 citations counted in INSPIRE as of 17 Sep 2013



\bibitem{atlasstop-c}
G.~Aad {\it et al.}  [ATLAS Collaboration], ATLAS-CONF-2013-068.

\bibitem{atlassbottom}
G.~Aad {\it et al.}  [ATLAS Collaboration],
  %``Search for direct third-generation squark pair production in final states with missing transverse momentum and two b-jets in sqrt{s}=8 TeV pp collisions with the ATLAS detector,''
  arXiv:1308.2631 [hep-ex].
  %%CITATION = ARXIV:1308.2631;%%
  %2 citations counted in INSPIRE as of 17 Sep 2013


\bibitem{cmsgaugino}
S.~Chatrchyan {\it et al.}  [CMS Collaboration], CMS PAS SUS-12-022.


\bibitem{hv}
C.~Han, X.~Ji, L.~Wu, P.~Wu and J.~M.~Yang,
  %``Higgs pair production with SUSY QCD correction: revisited under current experimental constraints,''
  arXiv:1307.3790 [hep-ph];
  %%CITATION = ARXIV:1307.3790;%%
  %1 citations counted in INSPIRE as of 17 Sep 2013
Eric Brownson, Nathaniel Craig, Ulrich Heintz, Gena Kukartsev, Meenakshi Narain, and Neeti
Parashar, presentation in Snowmass meeting, Minneapolis, 2013;
B.~Coleppa, F.~Kling and S.~Su,
  %``Exotic Higgs Decay via AZ/HZ Channel: a Snowmass Whitepaper,''
  arXiv:1308.6201 [hep-ph].
  %%CITATION = ARXIV:1308.6201;%%

%%%%%%%%%%%%%%%%%%%%%%%%%%%%%%%%%%%%%%%%%%%%%%%%%%%%%%%%%%%%%%%%%%%%%%%%%%%%%%%%



\end{thebibliography}
\end{document}